%% file: main.tex
\documentclass[sigconf]{acmart}

\renewcommand\footnotetextcopyrightpermission[1]{}

\AtBeginDocument{%
  }

\usepackage{tikz}
\usepackage{amsmath}
\usepackage{cleveref}
\usepackage{xurl}
\usepackage{multirow}
\usepackage{makecell}
\usepackage{framed} 
\usepackage{pifont}
\usepackage{algorithm}
\usepackage{algpseudocode}
\usepackage{subcaption}
\usepackage[most]{tcolorbox}

\makeatletter
\def\@authornotemark{}
\makeatother

\renewcommand{\thefootnote}{\fnsymbol{footnote}}

\author{%
Hyungyo Kim$^{1}$\footnotemark[2],
Nicholas Satchanov$^{1}$,
Hrishi Shah$^{1}$,
Gaohan Ye$^{1}$,
Jiaqi Lou$^{1}$,\\
Robert Walkup$^{2}$,
Shweta Salaria$^{2}$,
I-Hsin Chung$^{2}$,
Hubertus Franke$^{2}$, \\
Seetharami Seelam$^{2}$,
Apoorve Mohan$^{2}$\footnotemark[1],
Nam Sung Kim$^{1}$\footnotemark[1]
\vspace{6pt}
}

\affiliation{%
  \institution{\Large
  $^{1}$University of Illinois at Urbana-Champaign \qquad
  $^{2}$IBM Research}
  \country{}
  \vspace{6pt}
}

\renewcommand{\shortauthors}{Kim et al.}

\newcommand{\niparagraph}[1]{\noindent\textbf{#1}}
\newcommand{\ie}{\textit{i.e., }}
\newcommand{\eg}{\textit{e.g., }}

\newcommand{\framework}{\texttt{TRANSIT}}

\newtcolorbox{observationbox}[1]{
    colback=gray!6,
    colframe=gray!55,
    boxrule=0.5pt,
    arc=1pt,
    left=5pt,
    right=5pt,
    top=4pt,
    bottom=4pt,
    before skip=6pt,
    after skip=6pt,
    fonttitle=\bfseries,
    title={#1}
}

\begin{document}

\title{TRANSIT: Transparent Scale-in for Multi-Node LLM Training}



\keywords{}




\begin{sloppypar}

\input{Sections/0_abstract}

\maketitle

\footnotetext[2]{%
This work was initiated during Hyungyo Kim’s internship at IBM Research and was subsequently continued through the IBM-Illinois Discovery Accelerator Institute (IIDAI).
}

\footnotetext[1]{%
Corresponding authors: Apoorve Mohan
(Apoorve.Mohan@ibm.com) and Nam Sung Kim
(nam.sung.kim@gmail.com).
}

\renewcommand{\thefootnote}{\arabic{footnote}}

\pagestyle{plain}

\input{Sections/1_introduction}
\input{Sections/2_background}
\input{Sections/3_insights}
\input{Sections/4_proposed}
\input{Sections/5_results}
\input{Sections/7_related}
\input{Sections/8_conclusion}
\input{Sections/9_Acknowledgement}

\bibliographystyle{ACM-Reference-Format}
\bibliography{Sections/ref}

\end{sloppypar}
\end{document}

%% file: Sections/0_abstract.tex
\begin{abstract}

The growing memory requirements of large language models (LLMs) increasingly force distributed training jobs to span multiple GPU nodes to accommodate their training state. This reduces per-GPU computational efficiency, increases network traffic, and exacerbates job queueing delays. To address these inefficiencies, we introduce \framework, a system that transparently enables efficient distributed LLM training on fewer GPUs through memory oversubscription. \framework\ optimizes CPU--GPU data movement during oversubscription by coordinating driver-managed page residency with execution-guided zero-copy access, exploiting recurring kernel execution patterns across training iterations.

We evaluate \framework\ on dense and mixture-of-experts (MoE) models across different parallelization strategies using up to 64 NVIDIA H100-80GB GPUs interconnected over RDMA over Converged Ethernet (RoCE). Compared with the minimum GPU count required to accommodate the training state in GPU memory, \framework\ enables training on up to 50$\%$ fewer GPUs while retaining over 90$\%$ of the baseline per-GPU computational throughput. In some configurations, training on 17–33$\%$ fewer GPUs even improves throughput due to reduced distributed communication overhead. When training on fewer GPUs, \framework\ achieves up to 243$\%$ and 87$\%$ higher throughput than conventional driver-runtime-managed memory oversubscription and framework-managed CPU offloading, respectively, while reducing CPU memory consumption by up to 79$\%$ compared with the latter. Training on fewer GPUs with \framework\ also reduces per-node network traffic by up to 33$\%$ and improves throughput by up to 44$\%$ under constrained network bandwidth. Production-trace-driven simulations further show 15--49$\%$ lower inter-rack network traffic and up to 2--95$\%$ shorter job queue times for different cluster capacities.

\end{abstract}

%% file: Sections/1_introduction.tex
\section{Introduction}
\label{sec:intro}


\textbf{Forced Memory-Driven Scale-out.} The cost of training large language models (LLMs) has grown to the point where it shapes fundamental decisions about computing infrastructure and resource provisioning~\cite{brodsky2024hiddencosts}. For many training configurations, GPU allocations depend not only on computational requirements but also on the memory capacity needed to store parameters, gradients, optimizer states, and activations~\cite{gholami2024ai}. These demands can exceed the aggregate capacity of GPU high-bandwidth memory (HBM) available on a modest GPU allocation. For example, training Llama4-Scout~\cite{meta_llama4_2025} in our evaluated configuration (\S\ref{sec:experimental-setup}) requires 64 NVIDIA H100-80GB GPUs to accommodate its training state.

\textbf{Scale-out incurs well-known inefficiencies}. For example, it limits \texttt{(I1)}~\textit{training feasibility}: academic labs, startups, and enterprises operating under budget or infrastructure constraints may be unable to provision the required GPUs, while even well-funded organizations face availability, procurement, and supply-chain constraints~\cite{cheng2023towards}. It also introduces \texttt{(I2)}~\textit{computational inefficiency}: increasing model parallelism to accommodate training state can partition computation into smaller operations and add communication and synchronization overhead, yielding disproportionately small throughput gains from additional GPUs~\cite{fernandez2024hardware}. At the datacenter level, larger distributed configurations increase \texttt{(I3)}~\textit{network sensitivity} to bandwidth limitations, contention, and delays~\cite{si2025collective, cao2024crux, ghorbani2025congestion, zhao2024adapcc, rajasekaran2024cassini, shah2023taccl}; expand the \texttt{(I4)}~\textit{failure domain}, increasing exposure to GPU, node, and interconnect failures~\cite{kokolis2025revisiting}; and exacerbate \texttt{(I5)}~\textit{cluster resource inefficiency}, as fixed-size GPU allocations can experience substantial queueing delays under resource fragmentation and capacity constraints~\cite{peng2018optimus, 227623, xiao2018gandiva, hu2024characterization}.

\textbf{These inefficiencies motivate ``scale-in''}, \ie enabling distributed training jobs to execute on fewer GPUs than their memory requirements would otherwise dictate. By reducing memory-driven GPU allocations, scale-in can improve GPU resource efficiency, reduce communication demand, increase cluster-scheduling flexibility, and improve reliability from smaller distributed configurations.

\textbf{Challenge: Oversubscription Without Losing Throughput.}
Scale-in requires accommodating training state that exceeds the reduced aggregate GPU HBM capacity. Modern GPU servers offer an opportunity through substantial CPU DRAM capacity, which can serve as an additional memory tier. However, CPU DRAM has substantially lower bandwidth and higher access latency than HBM, and inefficient data movement over PCIe can impact training performance. The central challenge is therefore to oversubscribe GPU memory using CPU DRAM while preserving per-GPU training throughput; otherwise, memory-management overheads can negate the benefits of scale-in.

A major source of overhead is the CPU--GPU data movement during oversubscription. Driver-runtime-managed oversubscription (such as CUDA Unified Virtual Memory (UVM)~\cite{nvidia-uvm} and HIP Managed Memory~\cite{hip-managed-memory}) transparently migrates pages on-demand without application-level knowledge of access and reuse patterns, incurring page-fault overheads and repeated migration--eviction cycles that amplify PCIe traffic. These costs are high enough that PyTorch's documentation discourages using UVM for deep learning workloads~\cite{cuda-semantics}. Framework-managed offloading (such as implemented by DeepSpeed~\cite{ren2021zero, rajbhandari2020zero, 10.1145/3458817.3476205}, TorchTitan~\cite{liang2025torchtitan}, and Megatron-LM~\cite{shoeybi2019megatron, narayanan2021efficient}) leverages application-level knowledge to orchestrate tensor-level transfers, but shifts memory-management complexity to developers, requires framework-specific integration, and can incur substantial PCIe traffic-related overheads. 
Importantly, despite their different memory-management strategies, both approaches can incur substantial CPU--GPU data movement overhead, either through repeated page migrations under memory pressure or through coarse-grained tensor transfers. 
In our experiments, we observed that these memory oversubscription approaches reduce per-GPU training throughput by up to about $38\%$ when scaling Llama3-70B training from 24 to 16 NVIDIA H100-80GB GPUs (see \S\ref{sec:oversub-performance}).

The underlying challenge, thus, is balancing the benefits of HBM residency against the migration overhead during a training iteration. Migrating CPU-resident data into HBM incurs a one-time transfer cost that subsequent accesses amortize through HBM's higher bandwidth and lower latency; however, it consumes limited capacity and can evict useful pages. Conversely, direct CPU-memory access avoids migration but pays higher latency and PCIe cost on every access. Existing approaches~\cite{jung2023deepum, lin2025forest} primarily optimize which data to move and when, rather than avoiding migration through execution-guided direct access. This raises the following \textbf{research questions:} \emph{- Can recurring kernel execution provide sufficient guidance to decide when to enable zero-copy access to CPU-resident data rather than relying on page migration? - Can such an approach mitigate memory oversubscription overhead without explicit data placement, prefetching, or fine-grained memory-access tracking? - Can efficient memory oversubscription enable distributed LLM training on fewer GPUs while preserving high per-GPU computational efficiency?}


\textbf{Our Approach.} We present \textbf{\framework} (TRANsparent Scale-In for multi-node Training), a transparent GPU memory oversubscription system that coordinates driver-managed page residency with execution-guided zero-copy access to CPU-resident data, enabling efficient scale-in for distributed LLM training. By leveraging CPU DRAM as an extension of GPU HBM, \framework\ reduces memory-driven GPU requirements while preserving per-GPU training throughput. Unlike approaches that use tensor lifetimes~\cite{zhang2023g10, NEURIPS2025_46863e01} or page-access correlations~\cite{jung2023deepum} to predict and schedule data movement, \framework\ exploits recurring kernel-launch patterns to guide zero-copy activation without tracking individual memory accesses. The GPU driver manages the page residency, while \framework\ uses coarse-grained execution progress to decide when to enable zero-copy of CPU-resident pages. This allows efficient memory oversubscription without tensor-level knowledge, page-access profiling, explicit data-placement decisions, or framework integration. 

\textbf{Observations.} \framework's design is motivated by three observations about the limitations of migration-centric oversubscription and the tradeoffs of direct CPU-memory access (see \S\ref{sec:characterization} for details). \texttt{(O1)}~First, optimizing fine-grained migrations improves PCIe transfer efficiency but does not eliminate the cost of moving CPU-resident pages into HBM. \texttt{(O2)}~Second, under memory pressure, repeated migration and eviction can generate substantial data movement, which zero-copy access can avoid by accessing CPU-resident pages directly. \texttt{(O3)}~Third, naive zero-copy can significantly degrade performance because frequently accessed data benefits from HBM residency. However, we find that a broad range of zero-copy activation points preserves performance during training iterations, suggesting that recurring kernel-launch progress can guide when to enable zero-copy without tracking individual memory accesses or reconstructing tensor lifetimes.

\textbf{Design.} These observations show that the timing of zero-copy activation determines whether direct CPU-memory access complements or undermines the GPU driver's residency decisions. Before activation, on-demand migration lets frequently accessed pages (\ie hot pages) establish HBM residency; after activation, the GPU can access CPU-resident pages (\ie cold pages) directly through its L2 cache without migrating them into HBM. \framework\ realizes this with three user-space components (\Cref{fig:uvm_shim}): an allocation interceptor that redirects GPU allocations to managed memory, an activity recorder that counts kernel launches per iteration during warm-up, and a zero-copy controller that enables direct access for the relevant managed-memory regions once kernel-launch progress reaches the derived activation threshold (see \S\ref{sec:proposed} for details). The advice is applied once: pages resident in HBM at activation stay local, and pages in CPU memory are thereafter accessed in place rather than migrated. 
Activation thus preserves the existing access paths for HBM-resident pages while enabling direct access to CPU-resident pages, and the kernel-launch threshold determines when these mappings are established.
Each GPU acts independently, adding no inter-GPU communication or synchronization, and \framework\ requires no modifications to training applications, frameworks, communication libraries, schedulers, or device drivers.

This paper makes the following \textbf{contributions}:

\texttt{(C1)} We characterize the limitations of GPU memory oversubscription for distributed LLM training, demonstrating that improving fine-grained migration efficiency alone is insufficient under memory pressure. We quantify the costs of migration-induced data movement and the performance sensitivity of zero-copy activation timing, establishing the need to coordinate HBM residency with direct CPU-memory access.

\texttt{(C2)} We design and implement \framework, a lightweight GPU memory oversubscription system that exploits recurring kernel-launch patterns to guide zero-copy activation while retaining driver-managed page residency, without tensor-lifetime analysis, page-access profiling, or explicit data-placement decisions. Implemented as a user-space shared library, \framework\ enables transparent scale-in across training configurations without framework-managed offloading or modifications to the training stack.

\texttt{(C3)} We evaluate \framework\ on dense and mixture-of-experts LLMs across multiple parallelism configurations using up to 64 NVIDIA H100 GPUs interconnected over RoCE. Our evaluation shows up to $87\%$ higher per-GPU throughput than state-of-the-art framework-managed approaches, and up to 243\% higher per-GPU throughput than conventional driver-runtime-managed oversubscription (CUDA UVM). \framework\ enables training with up to $50\%$ fewer GPUs while maintaining over $90\%$ of baseline per-GPU throughput in the evaluated configurations. Scale-in reduces per-node network traffic by up to $33\%$ and improves per-GPU training throughput by up to $44\%$ in network-constrained environments. Production-trace-driven simulations further demonstrate reductions of 15--49$\%$ in inter-rack network traffic and up to 2--95$\%$ in job queueing time for different cluster capacities.

%% file: Sections/2_background.tex
\section{The Case for Scale-In}
\label{sec:motivation}


Distributed LLM training must balance computational efficiency, memory capacity, and communication overhead across GPUs and nodes. However, the GPU allocation needed to accommodate training state may differ from what efficient computation requires, creating opportunities for scale-in. This section reviews the factors behind memory-driven scale-out, its inefficiencies, and the conditions that motivate scale-in.

\subsection{Memory-Driven Scale-Out and Its Costs}
\label{sec:memory-scaleout}

\noindent\textbf{Memory requirements and parallelism.} GPU memory must accommodate model parameters, gradients, optimizer states, and activations during LLM training~\cite{rajbhandari2020zero}. Mixed-precision training~\cite{micikevicius2018mixed}, activation checkpointing~\cite{chen2016training}, and memory-efficient attention such as FlashAttention~\cite{dao2022flashattention} reduce this footprint, yet substantial memory requirements remain. Our Llama4-Scout~\cite{meta_llama4_2025} configuration (global batch size 64, sequence length 8192) requires about 4.3\,TB of aggregate GPU memory even with these optimizations, and therefore 64 NVIDIA H100-80GB GPUs in our evaluated configuration without offloading. Training frameworks distribute this state by combining tensor parallelism (TP)~\cite{shoeybi2019megatron}, pipeline parallelism (PP)~\cite{NEURIPS2019_093f65e0,10.1145/3341301.3359646}, fully sharded data parallelism (FSDP)~\cite{liang2025torchtitan} or ZeRO~\cite{rajbhandari2020zero}, and, for mixture-of-experts (MoE) models, expert parallelism (EP)~\cite{lepikhin2020gshardscalinggiantmodels,wang2025wlb}. Because partitioning training state also determines how computation and communication are distributed, increasing parallelism to meet memory constraints can reduce local computation while adding communication and synchronization overhead. Memory capacity can thus impose a minimum GPU allocation that differs from the one that uses computational resources most efficiently.


\noindent\textbf{Computational inefficiency.} Memory-driven scale-out can reduce computational efficiency when additional GPUs are allocated primarily to accommodate training state rather than increase useful computation. Increasing the degree of model parallelism can partition computation into smaller operations, reducing local work while adding communication and synchronization overhead~\cite{fernandez2024hardware,zheng2022alpa}. As training spans multiple nodes, collective communication becomes increasingly dependent on inter-node bandwidth and network conditions~\cite{shoeybi2019megatron,narayanan2021efficient}. Consequently, additional GPUs provisioned to satisfy memory requirements may contribute disproportionately little to aggregate training throughput. The extent of this inefficiency depends on the workload and parallelization strategy, as increasing GPU count does not necessarily reduce computational efficiency in every configuration.

\noindent\textbf{Network sensitivity and stragglers.} Distributed LLM training relies on collective communication, including all-gather, reduce-scatter, and all-to-all operations, to exchange model states, gradients, and intermediate results. These operations introduce synchronization dependencies, making training performance sensitive to network bandwidth, congestion, and slow participants~\cite{wu2024falcon}. In shared datacenters, network oversubscription and contention can further reduce effective bandwidth and introduce performance variability~\cite{gangidi2024rdma,qian2024alibaba,si2025collective}. Prior studies of production AI training clusters report network contention affecting up to approximately 36\% of training jobs~\cite{cao2024crux}, with congestion-related slowdowns exceeding 30\% in some settings~\cite{rajasekaran2024cassini}. 
We observe a similar straggler effect in our experiments. When training Llama4-Scout on 64 NVIDIA H100 GPUs across 16 nodes, reducing a single node's NIC bandwidth from 400\,Gbps to 100\,Gbps decreases overall training throughput by 50\%: because collectives synchronize participants, the slowest node
bounds the job's progress. Larger configurations involve more nodes and links that can degrade, increasing exposure to stragglers.

\noindent\textbf{Cluster scheduling and reliability.} Memory-driven scale-out also affects resource allocation at the datacenter level. Distributed training jobs commonly request a fixed number of GPUs that must be allocated before execution begins~\cite{peng2018optimus,xiao2018gandiva}. Under capacity constraints, larger allocations can experience prolonged queueing delays. Even when sufficient aggregate GPU capacity is available, resource fragmentation and topology-aware placement requirements can prevent a scheduler from satisfying a job's allocation request~\cite{hu2024characterization}. This is particularly consequential in development and experimentation environments, where queueing delays can limit iteration speed. Larger distributed configurations also involve more GPUs, nodes, and interconnect components, expanding the failure domain and increasing exposure to component failures and recovery overheads~\cite{kokolis2025revisiting}. Thus, provisioning additional GPUs primarily for memory capacity can impose costs beyond the computational resources the job consumes, including network sensitivity, scheduling constraints, and reliability risks.


\subsection{Scale-In Opportunity}
\label{sec:scalein-opportunities}

\noindent\textbf{Reducing memory-driven GPU allocations.}
The inefficiencies of memory-driven scale-out motivate an alternative approach: reducing the number of GPUs required for training by leveraging CPU DRAM as an additional memory tier. By moving part of the training state outside GPU HBM, scale-in increases the effective memory capacity per allocated GPU without adding GPUs primarily for memory. At the job level, scale-in can make training feasible on infrastructure that cannot accommodate the original GPU allocation. At the datacenter level, smaller allocations can improve scheduling flexibility, reduce resource fragmentation, and allow more jobs to execute concurrently under capacity constraints. Reducing the number of participating GPUs and nodes can also lower communication demand, although the effect depends on the parallelization strategy and workload. In particular, reducing GPU count does not necessarily reduce the communication volume of every collective operation.

\begin{figure}[t]
    \centering
    \includegraphics[width=\columnwidth]{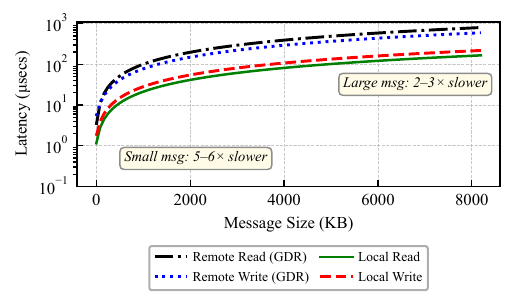}
    \caption{Read and write latencies for accessing local CPU DRAM over PCIe compared to remote GPU HBM accessed via GPUDirect RDMA (GDR) on NVIDIA H100 systems.}
    \label{fig:fetch_lat}
\end{figure}

\noindent\textbf{Local CPU DRAM versus remote GPU HBM.} When collectives span nodes, data movement depends on network bandwidth, latency, synchronization, and contention. In contrast, node-local CPU DRAM provides additional memory capacity without inter-node communication, albeit at lower bandwidth and higher latency than local GPU HBM. \Cref{fig:fetch_lat} compares local CPU DRAM access over PCIe Gen5 with remote GPU HBM access over a 400\,Gbps ConnectX-7 network using GPUDirect RDMA (GDR)~\cite{gdr,nvidia_cx}: local access achieves 5--6$\times$ lower latency for small transfers and 2--3$\times$ lower latency for larger transfers. These measurements highlight an opportunity for scale-in: local CPU DRAM can provide a competitive memory-access path relative to remote GPU transfers, enabling jobs to run on fewer nodes, potentially reducing distributed communication, and thus alleviating network pressure in bandwidth-constrained, congested, or oversubscribed environments.


\noindent\textbf{Performance and resource tradeoffs.} Scale-in consolidates training state and computation onto fewer GPUs, increasing per-device memory requirements. Using CPU DRAM as additional memory introduces access overhead relative to local HBM, while reducing GPU count can lower aggregate throughput and increase job completion time even when per-GPU efficiency is preserved. Nevertheless, smaller allocations can reduce queueing delays and network footprint while improving cluster scheduling flexibility. Importantly, fewer GPUs reduce total GPU-hours only if per-GPU throughput exceeds that of the original allocation. Thus, the objective is to reduce memory-driven GPU allocations while limiting the performance overhead of accessing training state outside HBM. We next examine GPU memory oversubscription mechanisms and LLM training memory-access patterns that determine this overhead.

\subsection{Memory Access and Management}
\label{sec:memory-oversubscription}


\begin{figure}[t]
    \centering
    \includegraphics[width=\linewidth]{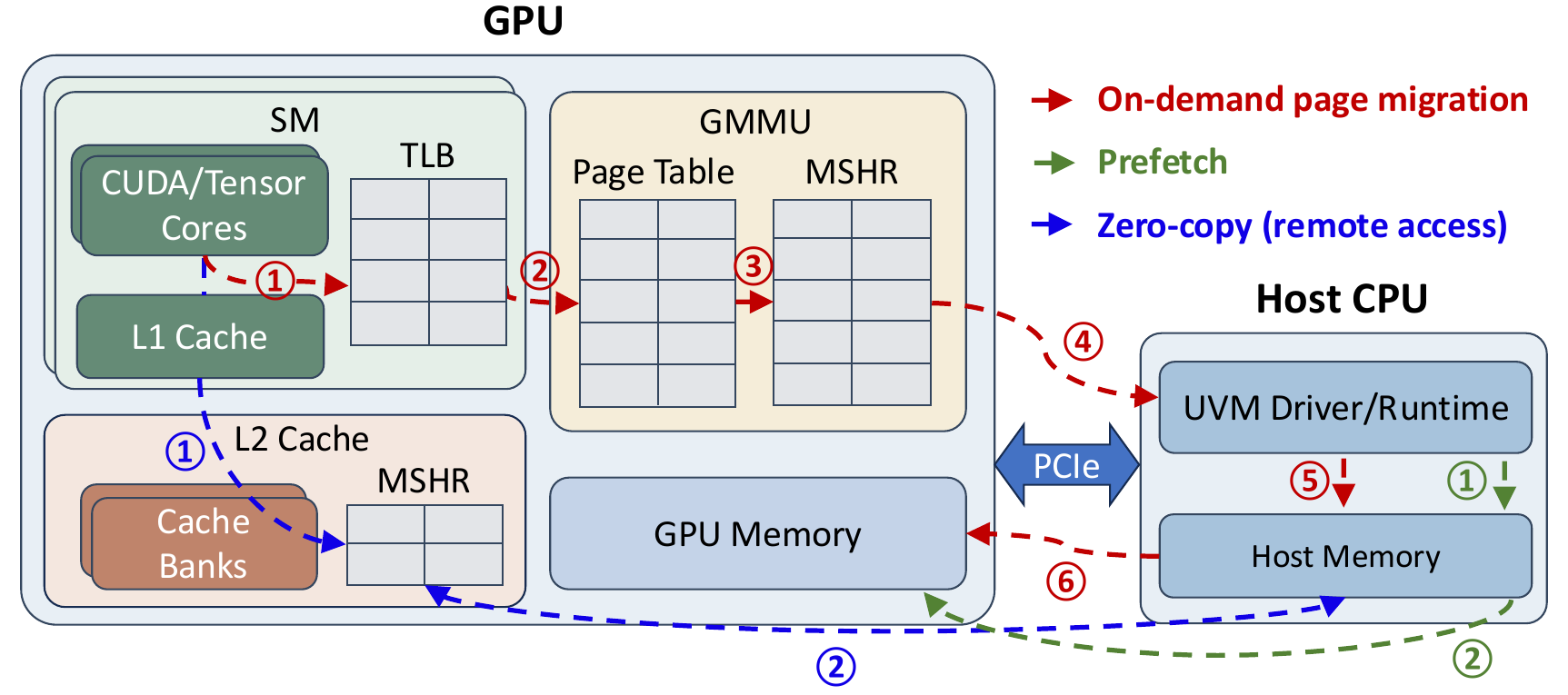}
    \caption{UVM data access paths: on-demand migration (red), prefetch (green), and zero-copy (blue).}
    \label{fig:uvm_background}
    \vspace{-12pt}
\end{figure}

\noindent\textbf{GPU memory oversubscription.} GPU memory oversubscription lets a workload's memory footprint exceed HBM capacity by leveraging larger-capacity CPU DRAM, whose access over PCIe incurs higher latency and lower bandwidth. Its efficiency depends on how data moves between these memory tiers and how often the GPU accesses CPU-resident data. Framework-managed offloading, as in DeepSpeed~\cite{ren2021zero, 10.1145/3458817.3476205} and TorchTitan~\cite{liang2025torchtitan}, places selected training state in CPU memory and exploits tensor lifetimes and training dependencies to schedule transfers and overlap them with computation. However, it requires framework-level coordination, typically manages transfers at tensor granularity, and its performance depends on transfer volume and computation overlap. In contrast, CUDA Unified Virtual Memory (UVM)~\cite{nvidia-uvm} provides driver-runtime-managed page residency, enabling transparent, page-granular oversubscription without explicit framework orchestration. Under demand-driven migration, GPU accesses to nonresident managed pages can trigger page faults and migrations from CPU DRAM into HBM (\Cref{fig:uvm_background}), incurring fault-handling and transfer overhead and potentially causing repeated migration--eviction cycles under HBM pressure. Alternatively, \emph{zero-copy} allows the GPU to access CPU-resident data directly over PCIe without first migrating it into HBM. While migration incurs an upfront transfer cost that can be amortized through subsequent HBM accesses, zero-copy avoids migration but incurs remote-memory access overhead. Their relative efficiency therefore depends on transfer overhead, data reuse, and HBM residency.

\begin{figure}[t]
    \centering
    \includegraphics[width=0.95\linewidth]{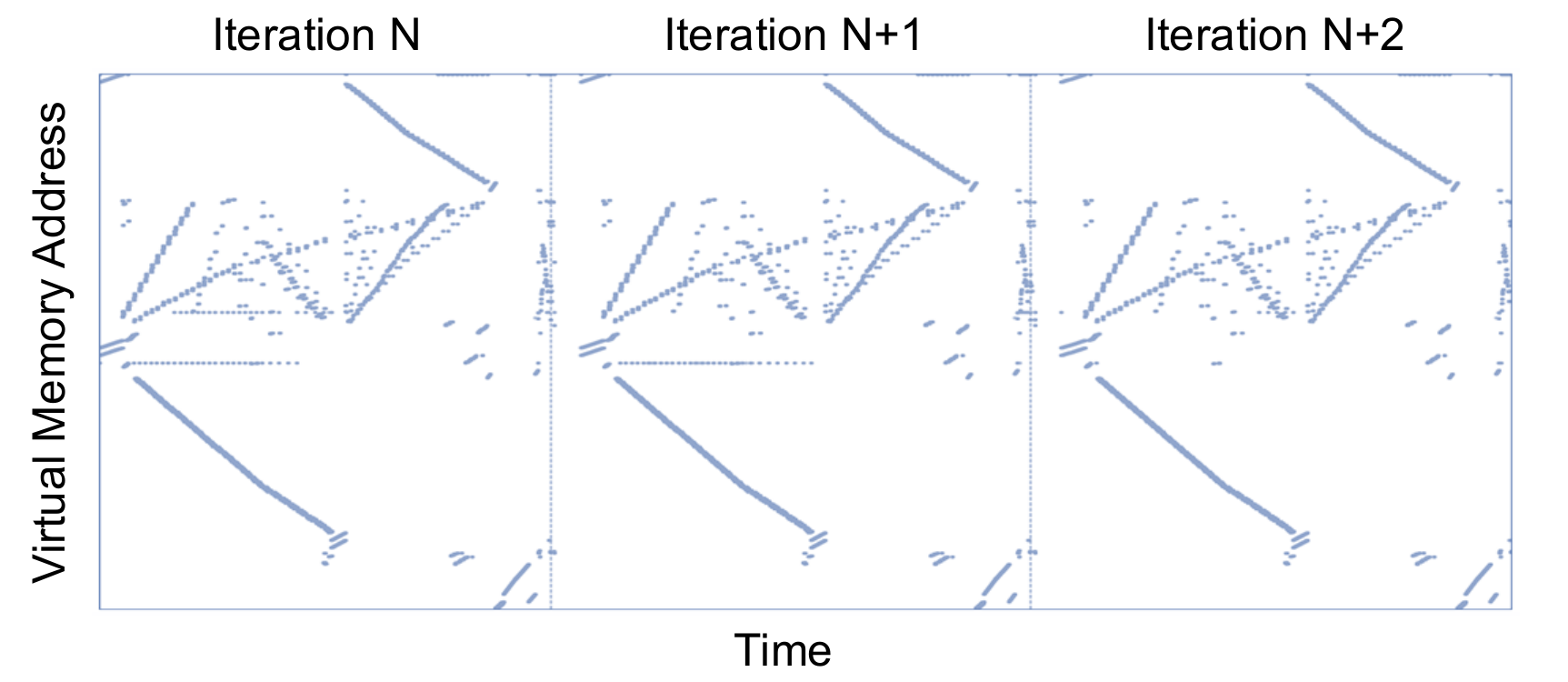}
    \caption{Data migration pattern during training of Llama3-8B on an NVIDIA H100 NVL (94\,GB HBM), with a memory footprint at 145\% of HBM capacity.}
    \label{fig:locality}
    \Description{Managed-memory migration addresses over three consecutive Llama3-8B training iterations.}
\end{figure}

\noindent\textbf{Memory-access and kernel-execution patterns.} LLM training exhibits structured spatial and temporal locality. Spatial locality arises from contiguous tensor accesses, while temporal locality reflects reuse of training state across execution phases. During the forward pass, parameters are accessed layer by layer to compute activations, which are reused during backward propagation. Parameters and gradients may be accessed repeatedly, particularly with gradient accumulation, whereas optimizer states and higher-precision parameter copies are accessed mainly during optimizer updates, resulting in longer reuse intervals. \Cref{fig:locality} illustrates these characteristics through UVM page-migration patterns over three iterations of Llama3-8B training on a single H100 NVL GPU, with a memory footprint at 145\% of HBM capacity. Contiguous migration regions reflect spatial locality, and phase-dependent migration activity indicates differences in temporal reuse. Because these execution phases exhibit recurring kernel-launch patterns across iterations, kernel-launch progress provides a coarse-grained indication of changing memory-access behavior without tracking individual accesses. Spatial locality favors bulk transfers, while temporal reuse affects whether migration costs can be amortized before eviction; exploiting both requires balancing HBM residency against repeated migration and direct CPU-memory access.


%% file: Sections/3_insights.tex
\section{Characterizing Memory Oversubscription}
\label{sec:characterization}

We present three key observations from characterizing driver-runtime-managed GPU memory oversubscription during LLM training: the limits of migration-efficiency optimization, the amplification of data movement under memory pressure, and the sensitivity of performance to zero-copy activation.

\subsection{Transfer Efficiency vs. Data Movement}
\label{sec:transfer-efficiency}

\begin{figure}[t]
    \centering
    \includegraphics[width=0.95\linewidth]{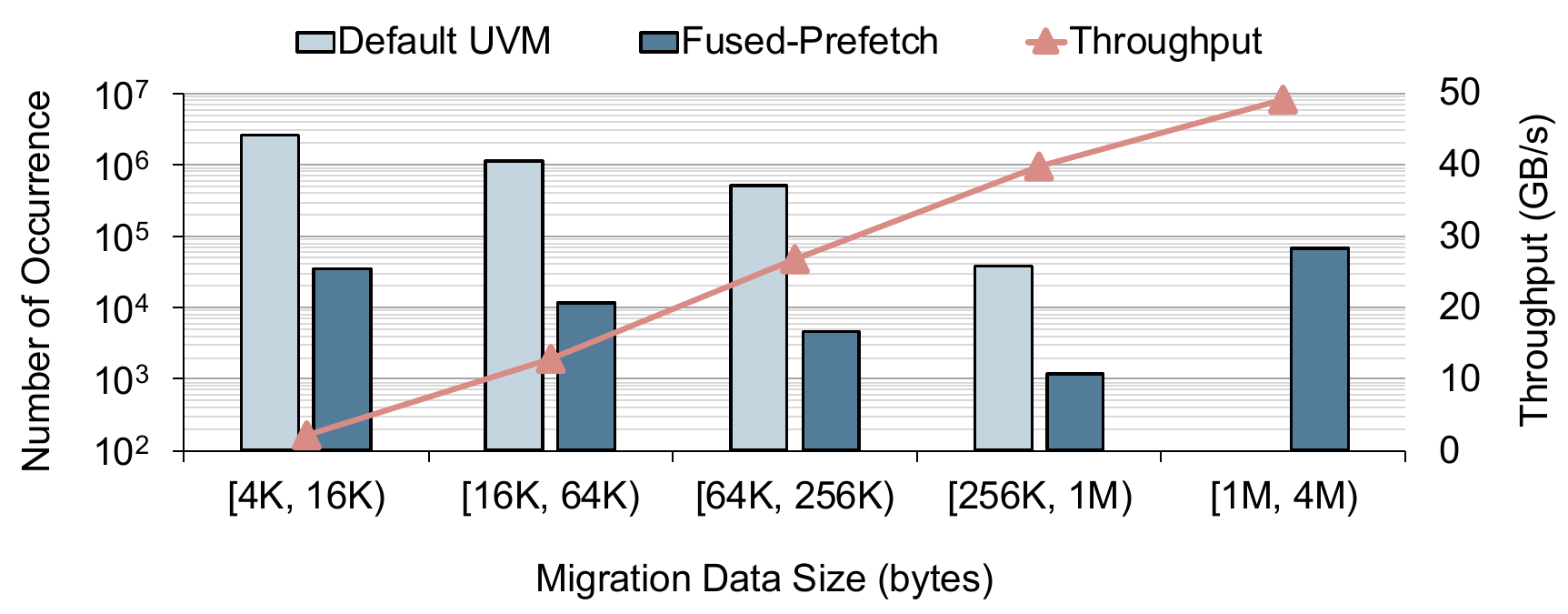}
    \caption{Number of occurrences and throughput of data migration events by migration size for UVM and Fused-Prefetch during Llama3-8B training on an H100 GPU.}
    \label{fig:page_mig_dist}
\end{figure}

\noindent\texttt{(O1) }\textbf{Optimizing Fine-Grained Migration Is Insufficient.} The repetitive execution and memory-access patterns of LLM training create opportunities to anticipate data accesses and proactively manage CPU--GPU transfers. We observe that conventional driver-runtime-managed oversubscription is dominated by small host-to-device migrations, typically ranging from a few KB to tens of KB, during Llama3-8B training on a single NVIDIA H100 GPU at 145$\%$ memory oversubscription (\Cref{fig:page_mig_dist}). Such transfers achieve low effective PCIe bandwidth because per-migration overheads are poorly
amortized. 

In contrast, device-to-host evictions occur in 2~MB units and achieve higher effective bandwidth. Exploiting the spatial locality of these accesses, we fuse small migration requests into larger asynchronous prefetch operations, termed \emph{Fused-Prefetch}, to improve transfer efficiency. This shifts the migration-size distribution toward transfers exceeding 1~MB (\Cref{fig:page_mig_dist}), reducing page faults by 97$\%$ and improving average PCIe bandwidth utilization by 2.4$\times$ compared with conventional demand-driven migration (\Cref{fig:uvm_stats}). Although fused prefetching improves effective PCIe bandwidth, it still relies on migrating CPU-resident pages into HBM rather than accessing them directly. Thus, improving transfer efficiency does not necessarily reduce the total volume of data migrated or the overhead of maintaining an oversubscribed HBM working set.

\subsection{Migration vs. Direct Access}
\label{sec:access-tradeoff}

\begin{figure}[t]
    \centering
    \includegraphics[width=0.95\columnwidth]{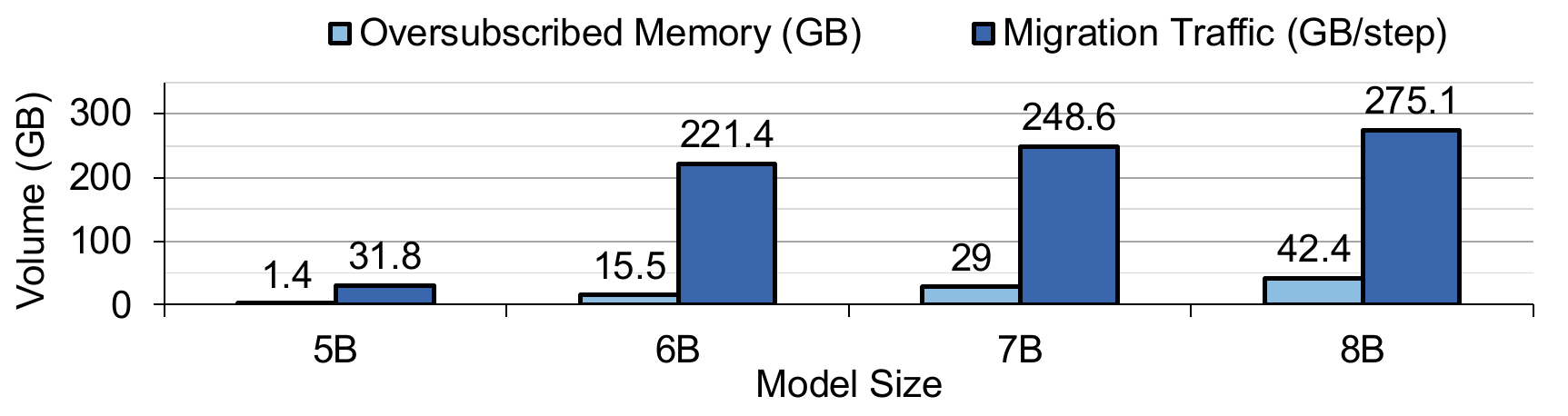}
    \caption{GPU memory oversubscription and migration traffic per training step for different Llama3-8B variant sizes.}
    \label{fig:migration_thrashing}
\end{figure}

\noindent\texttt{(O2) }\textbf{Faster Migration Does Not Prevent Thrashing.} Under HBM oversubscription, migrating CPU-resident pages into GPU memory can require evicting existing pages, causing working sets to compete for limited HBM capacity. Subsequent reuse of evicted pages can trigger further migrations, moving the same data repeatedly between CPU DRAM and HBM. Even modest oversubscription can therefore induce substantial data movement. \Cref{fig:migration_thrashing} shows Llama3-8B variants with 5B--8B parameters trained on a single H100 GPU with 94\,GB of HBM: although they exceed HBM capacity by only 1.4--42.4~GB, demand-driven migration generates bidirectional traffic equivalent to 6.5--22.7$\times$ the excess memory footprint per training step. For example, the 6B variant exceeds HBM capacity by just 15.5~GB but generates 221.4~GB of migration traffic per step, a 14.3$\times$ amplification. This behavior is particularly inefficient for training state with long reuse distances, where migration costs may not be amortized before eviction.

\begin{figure}[t]
    \centering
    \includegraphics[width=0.95\linewidth]{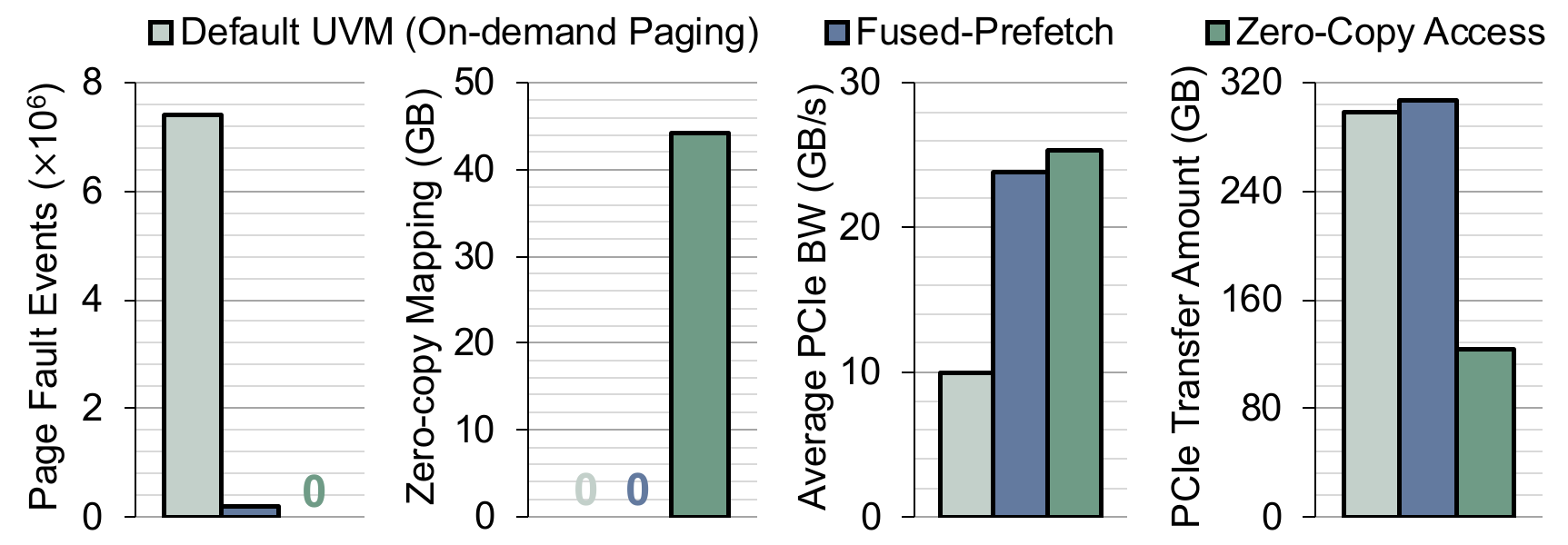}
    \caption{UVM and PCIe statistics of UVM, Fused-Prefetch, and Zero-Copy Access during Llama3-8B training on an H100 GPU with memory oversubscription. Page faults and PCIe transfer volume are measured over a single training iteration.}
    \label{fig:uvm_stats}
\end{figure}

Zero-copy provides an alternative access path: the GPU accesses CPU-resident pages directly over PCIe without first migrating them into HBM, avoiding HBM capacity consumption for those pages and the associated migration and eviction overheads, but incurring remote-memory access costs. For Llama3-8B training at 145$\%$ memory oversubscription, zero-copy reduces measured PCIe transfer volume by approximately $3\times$ and improves effective bandwidth utilization by $2.5\times$ compared with conventional demand-driven migration (\Cref{fig:uvm_stats}). Although zero-copy accesses still generate PCIe traffic, these results demonstrate that avoiding migration can reduce data movement beyond what more efficient migration alone can achieve. Its benefit, however, depends on data reuse: frequently reused data benefits from HBM residency, whereas data with limited reuse may be more efficiently accessed directly from CPU memory.

\subsection{Execution Progress and Memory Access}
\label{sec:execution-patterns}

\noindent\texttt{(O3) }\textbf{Naive Zero-Copy Degrades Performance.} LLM training exhibits heterogeneous memory-access patterns: frequently accessed data, such as activations and half-precision model parameters, benefits from HBM residency, whereas data with longer reuse intervals, including optimizer states and full-precision parameter copies, can be served directly from CPU DRAM. Forward propagation, backward propagation, and optimizer updates recur every iteration, and we observe recurring memory-migration patterns across iterations (\Cref{fig:locality}), suggesting that execution progress can indicate changes in memory-access behavior. To evaluate the sensitivity to activation timing, we examine how training throughput varies when zero-copy is enabled at different points within an iteration, expressed as a fraction of completed kernel launches (\S\ref{sec:sensitivity}). Enabling zero-copy too early can prevent frequently accessed pages from establishing HBM residency, whereas enabling it too late can incur unnecessary migration and eviction overhead. Poorly timed activation can degrade throughput by more than $50\times$, whereas activation within a broad window spanning approximately 40--90\% of kernel launches achieves comparable performance in the evaluated configuration.

This broad window suggests that recurring kernel-launch progress can provide a practical signal for guiding zero-copy activation without reconstructing tensor lifetimes, predicting page accesses, or explicitly classifying training state. Execution progress does not identify which pages reside in HBM or predict their future reuse; instead, it lets access-path selection align with existing driver-managed memory residency.

%% file: Sections/4_proposed.tex
\section{\framework: Overview}
\label{sec:proposed}

Building on the observations in \S\ref{sec:characterization}, efficient scale-in requires GPU memory oversubscription to preserve per-GPU training throughput on fewer GPUs; otherwise, savings in GPU allocation are offset by longer training times. We identify four requirements. \texttt{(R1)}~\textit{Efficient data movement}: the system should preserve HBM residency for frequently accessed data while avoiding migrations whose costs cannot be amortized by subsequent accesses.
\texttt{(R2)}~\textit{Low overhead}: memory-access-path selection should not require tensor-level memory management, page-access tracking, or significant runtime overhead. \texttt{(R3)}~\textit{Multi-node effectiveness}: benefits should hold across distributed training configurations, where overhead on individual GPUs can delay synchronized training steps, without introducing additional communication or synchronization. \texttt{(R4)}~\textit{Transparency}: the system should require no modifications to training applications, frameworks, communication libraries, schedulers, or device drivers.

\subsection{System Architecture}
\label{sec:architecture-workflow}

\begin{figure}[t]
    \centering
    \includegraphics[width=\linewidth]{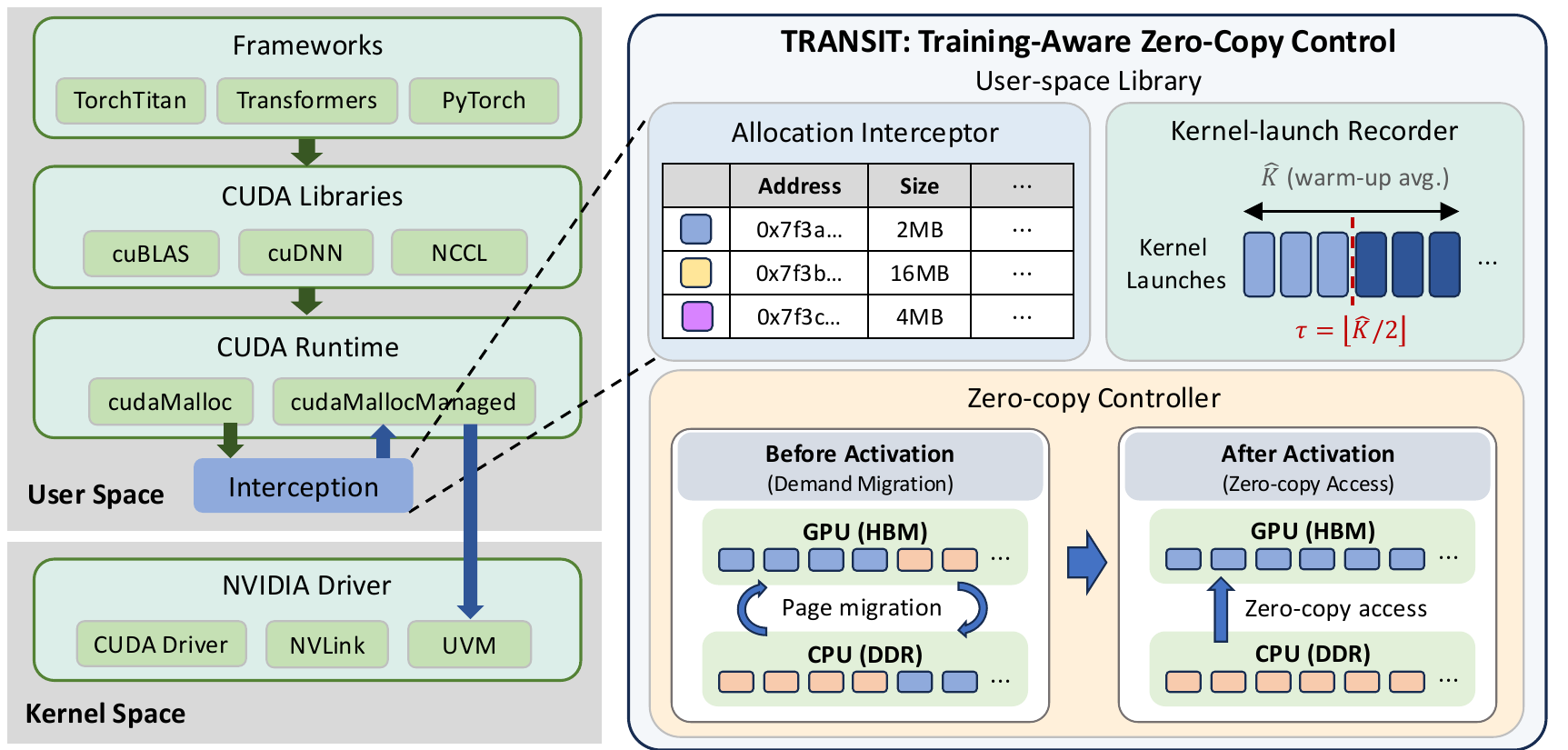}
    \caption{\framework\ architecture: a user-space allocation interceptor and activity recorder provide the managed-memory regions and execution progress used to coordinate access through CUDA runtime interfaces.}
    \label{fig:uvm_shim}
\end{figure}

\Cref{fig:uvm_shim} illustrates \framework's
architecture, which consists of three components within a user-space shared library, each maintained independently by every GPU worker process. The \emph{memory allocation interceptor} redirects supported GPU allocations to driver-runtime-managed memory (\eg CUDA Unified Virtual Memory~\cite{nvidia-uvm} or HIP Managed Memory~\cite{hip-managed-memory}) and records each allocation's base address and size in a process-local region table, without requiring tensor semantics or determining where individual pages reside. The \emph{kernel-launch activity recorder} identifies training-iteration boundaries and counts kernel launches per iteration during initial training iterations, providing a coarse-grained measure of execution progress at the cost of a counter update per launch. The \emph{zero-copy controller} uses these counts to set an activation threshold and, once it is reached, enables direct GPU access to CPU-resident pages in the recorded regions; its cost scales with the number of recorded regions rather than the number of pages. \framework\ thus proceeds in three stages: initialization, observation, and activation.

\textbf{Transparent Memory Oversubscription.} Managed allocations share a unified virtual address space accessible by the GPU, allowing training state to exceed HBM capacity without application changes. Before activation, the GPU driver and runtime manage residency unchanged: a GPU access to a nonresident page raises a page fault that the driver services by migrating the page from CPU DRAM into HBM, stalling the faulting access until migration completes. Under memory pressure, the driver may evict resident pages to CPU DRAM (\S\ref{sec:access-tradeoff}). During initial training execution, frequently accessed pages, such as activations and half-precision parameters, therefore become resident in HBM. In contrast, pages with longer reuse intervals, such as optimizer states and full-precision parameter copies, may be evicted as
the working set competes for limited HBM capacity. \framework\ preserves this driver-managed residency and only changes how the GPU subsequently accesses CPU-resident pages, reducing migration pressure without explicitly selecting which pages reside in HBM or CPU memory.

\textbf{Execution-Guided Activation.} \Cref{alg:transit-advise} summarizes the procedure, which measures kernel-launch counts during warm-up (Phase 1) and enables zero-copy access at the resulting threshold (Phase 2). As established in \S\ref{sec:execution-patterns}, LLM training exhibits recurring kernel-execution patterns across iterations. During the first $W$ iterations, the recorder counts the kernel launches $K_i$ in each iteration $i$, identifying iteration boundaries by intercepting \texttt{optimizer.step()} and observing launches through GPU runtime profiling callbacks, and computes $\widehat{K} = \frac{1}{W}\sum_{i=1}^{W} K_i$. We use $W=2$ across all evaluated workloads; these iterations both characterize execution progress and allow the memory manager to establish an initial HBM working set. The activation threshold is $\tau = \lfloor \widehat{K}/2 \rfloor$, the iteration midpoint, which lies within the broad favorable window identified in \S\ref{sec:execution-patterns}. Rather than identifying an exact transition between training phases, $\tau$ provides a workload-relative execution point determined through lightweight kernel-launch monitoring, without tracking individual memory accesses or reconstructing tensor lifetimes. We use launch counts rather than elapsed time because an iteration's kernel sequence is repeatable, whereas iteration time varies with communication delays and stragglers. Because kernel launches are asynchronous, host-side launch progress may run ahead of device execution; the breadth of the favorable window tolerates this offset. Each worker computes its own $\tau$, so ranks that execute different kernel sequences, such as pipeline stages, require no coordination.

\begin{algorithm}[t]
\caption{\framework: Execution-Guided Zero-Copy Activation}
\label{alg:transit-advise}
\begin{algorithmic}[1]
\Require GPU $d$, warm-up iterations $W$,
         managed-memory regions $\mathcal{R}$
\State $K_{\mathrm{sum}} \gets 0$
\Statex \textit{Phase 1: Measure kernel-launch counts}
\For{$i = 1$ to $W$}
    \State $k \gets 0$
    \For{each kernel launch in iteration $i$}
        \State $k \gets k + 1$
    \EndFor
    \State $K_{\mathrm{sum}} \gets K_{\mathrm{sum}} + k$
\EndFor
\State $\widehat{K} \gets K_{\mathrm{sum}}/W$
\State $\tau \gets \lfloor \widehat{K}/2 \rfloor$
\Statex \textit{Phase 2: Enable zero-copy access}
\State $k \gets 0$
\For{each kernel launch in the next iteration}
    \State $k \gets k + 1$
    \If{$k \geq \tau$}
        \For{each region $(p,n) \in \mathcal{R}$}
            \State $\Call{EnableZeroCopy}{p,n,d}$
        \EndFor
        \State \textbf{break}
    \EndIf
\EndFor
\end{algorithmic}
\end{algorithm}

In the next iteration, demand-driven migration continues until the launch count reaches $\tau$, allowing frequently accessed data to establish HBM residency before the access-path transition. At that point, the controller enables \emph{zero-copy access} to every recorded region. Zero-copy establishes GPU page-table mappings to CPU-resident pages without migrating their contents or allocating HBM. GPU loads and stores to these pages are then issued as fine-grained memory transactions over PCIe and served through the GPU's L2 cache (\Cref{fig:uvm_background}), rather than raising page faults that migrate entire pages into HBM. Pages already resident in HBM remain locally mapped and continue to be accessed at HBM bandwidth. After activation, CPU-resident pages no longer fault and are therefore not migrated into HBM, so steady-state residency is largely fixed at the activation point; the driver continues to manage eviction. Zero-copy thus trades a per-access interconnect cost for eliminating migration, eviction, and fault handling for CPU-resident data, addressing the tradeoff identified in \S\ref{sec:access-tradeoff}. 

\textbf{Steady-state memory access.} The mappings are established once and remain effective throughout subsequent training; \framework\ neither switches back to migration nor recomputes $\tau$, and kernel-launch monitoring is no longer required for access-path selection. Activation thus preserves the page residency, and $\tau$ determines its timing: by mid-iteration, the reused working set (parameters, gradients, and activations) is HBM-resident while once-per-step optimizer state has been evicted, and because iterations repeat, this residency state remains valid across subsequent iterations. \framework\ never identifies hot or cold pages or enforces their placement: the resulting behavior emerges from the memory manager's residency decisions before activation and the direct-access mappings established afterward.

\textbf{Memory consistency.} \framework\ changes only how the GPU maps managed pages: enabling direct access neither copies nor migrates data, and the underlying memory manager remains responsible for keeping mappings valid. \framework\ therefore inherits the memory-consistency semantics of driver-runtime-managed memory, introduces no separate coherence protocol, and does not alter the application's synchronization requirements; concurrent accesses to shared data follow the platform's standard synchronization and
memory-ordering rules.

\subsection{Implementation and Deployment}
\label{sec:implementation-deployment}

We implement \framework\ for NVIDIA GPUs as a user-space shared library that maps each design component onto CUDA Unified Virtual Memory (UVM) and the CUDA Profiling Tools Interface (CUPTI), requiring no modifications to the training stack.

\framework\ is loaded via \texttt{LD\_PRELOAD} and interposes on CUDA runtime allocation entry points (\eg \texttt{cudaMalloc}), forwarding all other calls to the original library. Each intercepted request is served by \texttt{cudaMallocManaged}, which reserves a range in the unified virtual address space without immediately backing it with HBM; physical pages are populated on first touch and thereafter migrated or evicted by the UVM driver at page granularity. Because frameworks such as PyTorch obtain device memory through a caching allocator that requests large
segments and sub-allocates tensors within them, interposing at this level
covers training state without framework changes and keeps the region table small.  The interceptor records each region's base address and size in a process-local table. Oversubscription requires a GPU and driver that support demand paging of managed memory, as on the evaluated H100 systems.

The recorder subscribes to CUPTI runtime and driver API callbacks for kernel-launch entry points and increments a process-local counter on each launch; it records no arguments, timestamps, or memory addresses, so the per-launch cost is a callback and a counter update. Iteration boundaries are identified by intercepting Python-level \texttt{optimizer.step()} calls through a runtime-installed hook, without modifying the training application, and the kernel-launch counter $K_i$ is recorded and reset at each boundary. Kernels launched as part of a CUDA Graph are not observed individually; the evaluated configurations do not use CUDA Graphs, and supporting them would require counting graph launches or the kernels within them. 

When the counter reaches $\tau$, the controller invokes \texttt{cudaMemAdvise} with \texttt{cudaMemAdviseSetAccessedBy} on each recorded region, specifying the local GPU. This advice directs the driver to keep the region mapped in the GPU's page tables regardless of where its pages reside: CPU-resident pages become directly accessible over PCIe without faulting, HBM-resident pages remain local, and if the driver later evicts a page to CPU memory, the GPU mapping is updated rather than removed, so subsequent accesses are served remotely instead of triggering re-migration~\cite{setaccessedby}. The controller issues no explicit migration, prefetch, or eviction commands, leaving residency decisions to UVM; activation costs one advice call per region, independent of the number of pages. \texttt{cudaMemAdviseSetAccessedBy} establishes mappings only; unlike \texttt{cudaMemAdviseSetReadMostly}, it creates no read-only replicas, so managed-memory consistency semantics are unchanged. HIP exposes analogous interfaces (\texttt{hipMallocManaged} and \texttt{hipMemAdviseSetAccessedBy}), so the design could be extended to AMD GPUs; we leave this port to future work.

\framework\ runs in multi-GPU and multi-node environments, including container-orchestrated clusters using Slurm, Kubernetes, and Red Hat OpenShift. Deployment requires only making the shared library available inside the training container and setting \texttt{LD\_PRELOAD} at job launch; evaluated configurations need no privileged containers, kernel modules, or modifications to the host operating system, GPU driver, or cluster scheduler. Because all \framework\ state is process-local, FSDP, TP, and PP retain their original communication and synchronization behavior, and NCCL runs unmodified.

%% file: Sections/5_results.tex
\section{Evaluation}
\label{sec:evaluation}
We evaluate \framework\ to answer four questions: (1) How effectively does training-aware zero-copy access mitigate the overhead of GPU memory oversubscription compared with existing offloading approaches? (2) How much can GPU allocation be reduced while preserving per-GPU training throughput? (3) How does scale-in affect performance under constrained inter-node bandwidth? (4) How do smaller GPU allocations translate into cluster-level scheduling and communication benefits?

\subsection{Experimental Setup}
\label{sec:experimental-setup}
\niparagraph{Training workload.} 
We use TorchTitan~\cite{liang2025torchtitan} as the training framework and evaluate four representative models: Llama3-70B~\cite{grattafiori2024llama3herdmodels}, Llama4-Scout~\cite{meta_llama4_2025}, Qwen3-32B~\cite{yang2025qwen3technicalreport}, and Qwen3-30B-A3B~\cite{yang2025qwen3technicalreport}, as summarized in Table~\ref{tab:models}.
Our evaluation covers fully sharded data parallelism (FSDP) and hybrid configurations that combine intra-node tensor parallelism (TP) with inter-node FSDP and pipeline parallelism (PP).
We use a local batch size of 1 for dense models and 4 for MoE models, increasing it for PP to mitigate pipeline bubbles.
To ensure compatibility across GPU counts with different data-parallel degrees, we set the global batch size to their common multiple.
We use TorchTitan's default mixed-precision training (FP32 parameters and BF16 computation), except for TP+PP configurations, which use BF16 throughout due to TorchTitan's limitations.
Full activation checkpointing is enabled across all configurations.

\niparagraph{Metrics and baselines.}
We report per-GPU TFLOPS, computed as floating-point operations per iteration divided by iteration time and GPU count; aggregate throughput is the product of per-GPU TFLOPS and GPU count.
We measure per-node network traffic using NVIDIA ConnectX RoCE hardware counters exposed through Linux sysfs, summing transmitted and received bytes per iteration.
We also report the CPU-resident memory footprint and PCIe traffic to characterize offloading overhead.

For each workload, the no-offloading baseline uses the minimum GPU count required without offloading.
At reduced GPU counts, we compare \framework\ against default UVM, Fused-Prefetch (\S\ref{sec:transfer-efficiency}), TorchTitan-Offload, ZeRO-Offload~\cite{ren2021zero}, and ZeRO-Infinity~\cite{10.1145/3458817.3476205}.
UVM uses default managed-memory behavior, while Fused-Prefetch coalesces prefetch transfers.
TorchTitan-Offload uses \texttt{enable\_cpu\_offloading=true}, while ZeRO-Offload and ZeRO-Infinity use their corresponding DeepSpeed configurations.

\begin{table}[t]
\footnotesize
\caption{Summary of the evaluated models, their architectural types, parallelization strategies, global and local batch sizes (GB and LB), and the corresponding minimum GPU count required without memory offloading.}
\label{tab:models}
\begin{tabular} {lrrrrr}
\hline
\toprule  
\textbf{Model} & \textbf{Type} & \textbf{Parallelism} & \textbf{GB} & \textbf{LB} & \textbf{\makecell[r]{Min.\ Num.\ \\ of GPUs}} \\ \midrule 
Qwen3-32B & Dense & FSDP & 128 & 1 & 8 \\ \cmidrule{1-6}
\multirow{3}{*}{Llama3-70B} & \multirow{3}{*}{Dense} & FSDP & 240 & 1 & 24 \\ 
 & & FSDP+TP & 48 & 1 & 16\\ 
 & & TP+PP & 30 & 30 & 12 \\ \cmidrule{1-6}
Qwen3-30B-A3B & MoE & FSDP & 960 & 4 & 20 \\ \cmidrule{1-6} 
Llama4-Scout & MoE & FSDP+TP & 64 & 4 & 64 \\ \bottomrule
\end{tabular}
\end{table}

\begin{table}[t]
\footnotesize
\centering
\caption{Summary of the multi-node cluster configurations used for the evaluation.} \label{tab:cluster_setup} \begin{tabular}{lcc} \toprule \textbf{Configuration} & \textbf{Cluster-A} & \textbf{Cluster-B} \\ \midrule 
Nodes & 16 & 3 \\ 
GPUs per node & 4$\times$ H100 (80GB) & 8$\times$ H100 (80GB) \\ 
CPU--GPU interconnect & PCIe Gen4 & PCIe Gen5 \\
NICs per node & 4$\times$ ConnectX-7 & 8$\times$ ConnectX-7 \\ 
NIC line rate & 400\,Gbps & 400\,Gbps \\ 
NVLink BW per GPU & 300\,GB/s & 900\,GB/s \\ \midrule
Host CPU & \multicolumn{2}{c}{Intel Sapphire Rapids, 80 cores} \\ 
CPU memory per node & \multicolumn{2}{c}{2\,TB} \\ \bottomrule \end{tabular} 
\end{table}

\niparagraph{Multi-node cluster.}
We evaluate \framework\ on two H100-based GPU clusters deployed in OpenShift, as summarized in \Cref{tab:cluster_setup}.
Cluster-A spans two racks with eight nodes each, interconnected through directly connected top-of-rack (ToR) switches providing 10.4\,Tbps of aggregate switching capacity.
Unless otherwise specified, all NICs operate at their nominal 400\,Gbps line rate.

We additionally evaluate bandwidth-limited configurations to capture settings where the available network bandwidth is lower than the nominal line rate.
Large-scale GPU clusters are increasingly exposed to partial hardware and network degradations, where failures or degraded components can reduce the throughput of otherwise healthy distributed training jobs~\cite{arfeen2025nonuniformtensorparallelismmitigatinggpufailure, kokolis2025revisiting}.
Beyond these degradations, bandwidth limitations also arise from deployment constraints, as academic or shared research clusters may not be provisioned with uniformly high-bandwidth interconnects.
To study such cases in a controlled setting, we emulate a degraded or bandwidth-limited scenario by capping the NICs to 100\,Gbps using \texttt{mlnx\_qos}.

\subsection{Memory Oversubscription Performance}
\label{sec:oversub-performance}
\begin{figure}[t]
    \centering
    \begin{subfigure}[b]{0.63\linewidth}
        \centering
        \includegraphics[width=\textwidth]{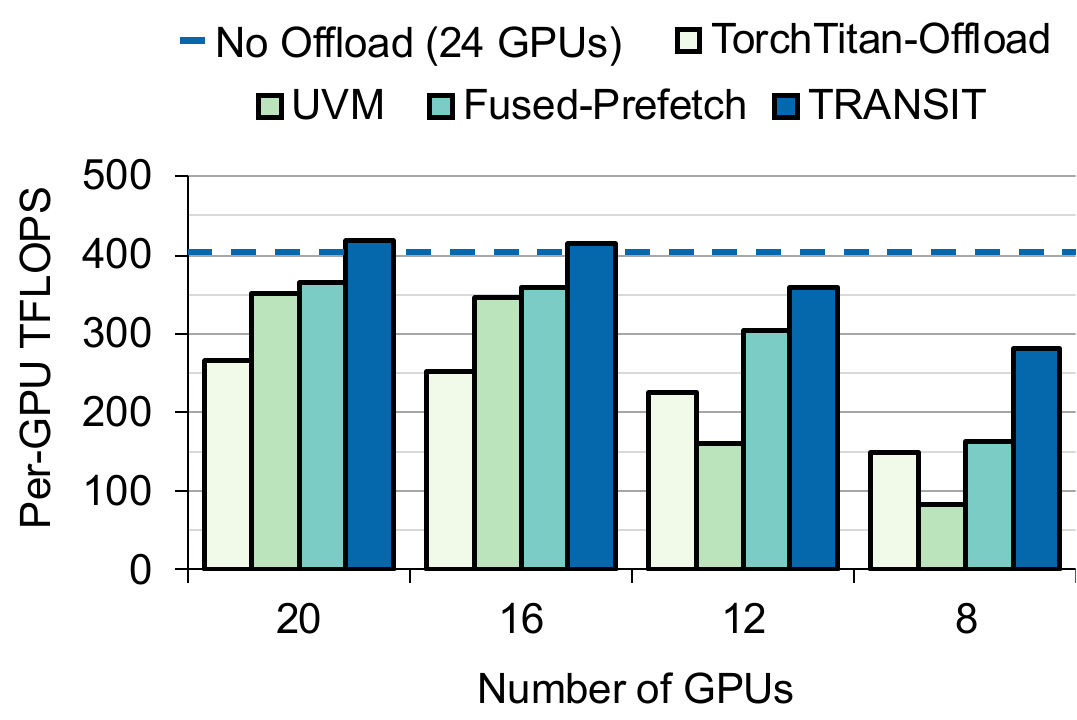}
    \end{subfigure}
    \hfill
    \begin{subfigure}[b]{0.36\linewidth}
        \centering
        \includegraphics[width=\textwidth]{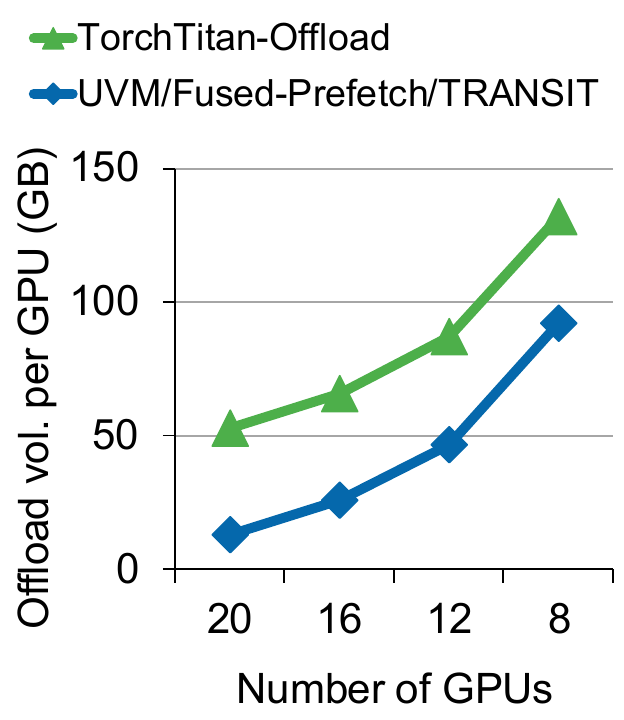}
    \end{subfigure}
    \caption{Per-GPU training throughput and CPU-resident memory footprint across GPU counts for Llama3-70B training with FSDP. The dashed line on the left panel indicates the 24-GPU no-offloading baseline.}
    \label{fig:result:scale_in}

\end{figure}

\niparagraph{Impact of memory access strategies.}
\Cref{fig:result:scale_in} compares Llama3-70B FSDP training on Cluster-A as the allocation shrinks below the 24-GPU, six-node no-offloading baseline.
At the same reduced GPU counts, both Fused-Prefetch and \framework\ outperform UVM, improving per-GPU TFLOPS by 4--99\% and 10--243\%, respectively.
While Fused-Prefetch improves transfer efficiency, it retains migration and eviction overheads under HBM pressure, whereas \framework\ mitigates migration thrashing through zero-copy access.
Notably, with 17--33\% fewer GPUs, \framework\ slightly exceeds the no-offloading baseline by 3--4\%, as reduced communication volume outweighs CPU-memory access overhead.
As the GPU count decreases further, each GPU holds a larger shard of the training state, increasing CPU-memory accesses and reducing per-GPU throughput.
Nevertheless, \framework\ consistently outperforms TorchTitan-Offload, maintaining 57--87\% higher per-GPU TFLOPS at aggressive scale-in, where UVM and Fused-Prefetch fall below TorchTitan-Offload.
As shown in the right panel of \Cref{fig:result:scale_in}, all three UVM-based approaches maintain substantially smaller CPU-resident memory footprints than TorchTitan-Offload, as they spill only the memory exceeding HBM capacity rather than entire classes of training state.

\begin{figure}
    \centering
    \includegraphics[width=0.95\linewidth]{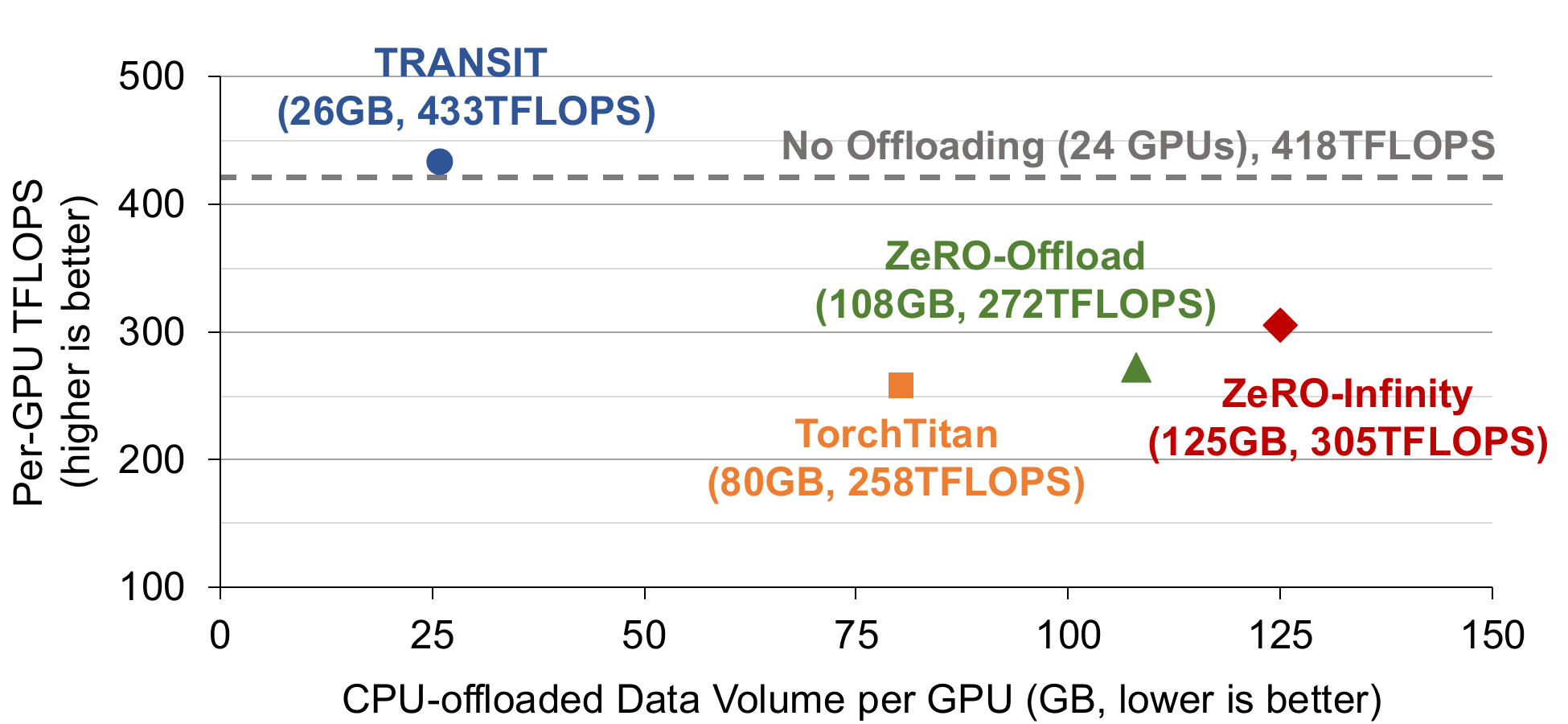}
    \caption{Per-GPU training throughput and offloaded data volume of \framework\ and framework-managed offloading approaches for Llama3-70B training on 16 GPUs. The dashed horizontal line indicates the 24-GPU no-offloading baseline.}
    \label{fig:result:perf_comparison}

\end{figure}

\niparagraph{Framework-managed offloading.}
Figure~\ref{fig:result:perf_comparison} compares offloading mechanisms for Llama3-70B on 16 GPUs across two Cluster-B nodes. The no-offloading baseline requires 24 GPUs and achieves 418 TFLOPS. At 16 GPUs, \framework\ achieves 433 TFLOPS, surpassing the baseline in per-GPU efficiency as reduced inter-node communication outweighs the overhead of CPU-memory accesses. In contrast, TorchTitan-Offload, ZeRO-Offload, and ZeRO-Infinity achieve 258, 272, and 305 TFLOPS, respectively. \framework\ thus delivers 68\%, 59\%, and 42\% higher per-GPU throughput than these framework-managed approaches, respectively.
Two differences in memory placement and access account for the observed performance gap. First, these baselines offload entire classes of training state regardless of how much exceeds HBM capacity, placing 80, 108, and 125\,GB per GPU in CPU memory, whereas \framework\ spills only the overflow (26\,GB, or 3.1--4.8$\times$ less) because driver-managed placement fills HBM before spilling.
Because these frameworks expose offloading only at the granularity of state classes, users cannot size the spill to the actual overflow; \framework\ does so automatically, as the overflow changes with model, parallelism, and GPU count.
Second, \framework\ accesses spilled state in place via zero-copy, whereas ZeRO-Offload and ZeRO-Infinity move it to host CPU cores for the optimizer update; a small spill alone is insufficient, as UVM and Fused-Prefetch spill the same volume yet underperform \framework\ (\Cref{fig:result:scale_in}).


\subsection{Scale-In Efficiency}
\label{sec:result:perf_scale_in}

\begin{figure}
    \centering
    \includegraphics[width=0.95\linewidth]{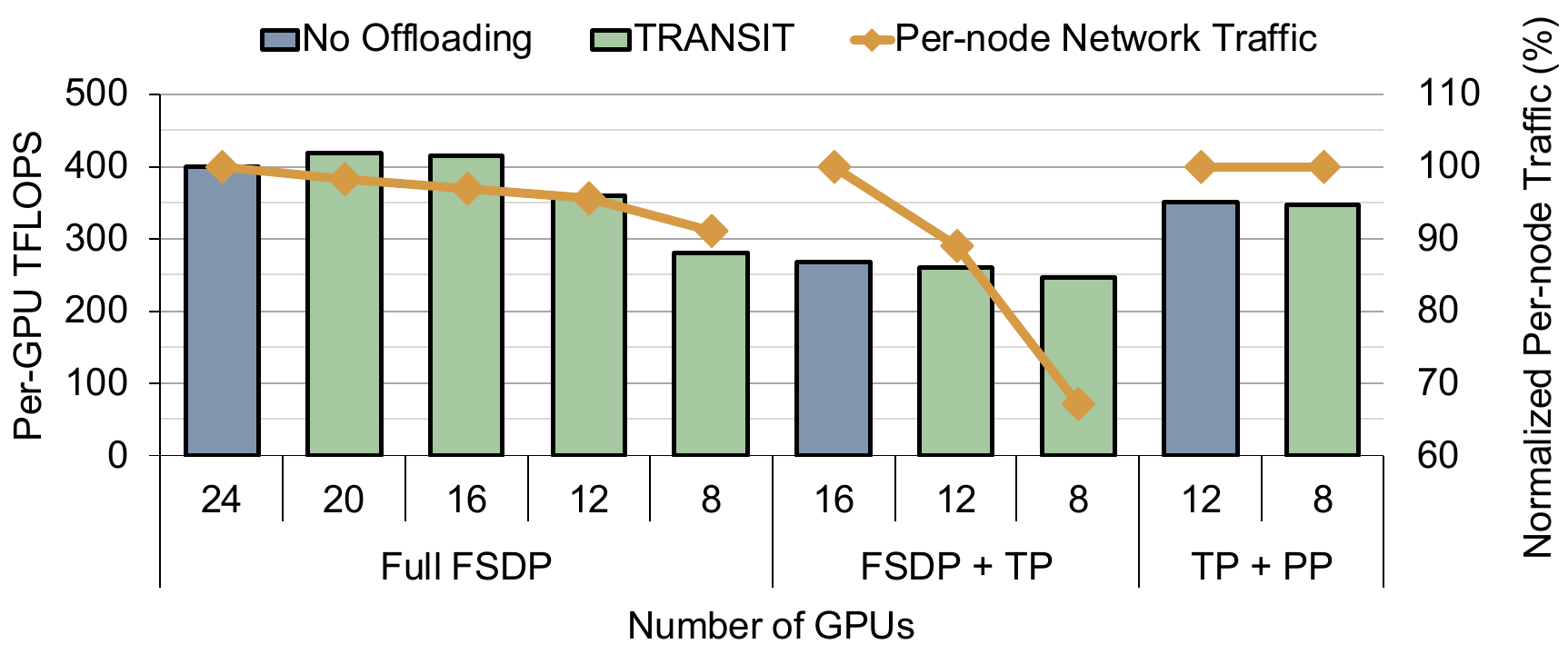}
    \caption{Per-GPU TFLOPS and normalized per-node network traffic for Llama3-70B training across parallelism strategies and GPU counts.}
    \label{fig:result:different_parallelism}
\end{figure}

\niparagraph{Different parallelisms.}
\Cref{fig:result:different_parallelism} shows the scaling behavior of Llama3-70B training on Cluster-A across different parallelization strategies (FSDP, FSDP+TP, and TP+PP), along with the corresponding per-node communication traffic as the number of nodes decreases.
Across all parallelization strategies, \framework\ maintains over 90\% of the per-GPU throughput at up to half the number of GPUs, relative to the no-offloading baseline.
For FSDP and FSDP+TP, scale-in also reduces per-node communication volume per step.
This trend arises because FSDP’s collective operations (all-gather and reduce-scatter) scale with model sharding, causing communication volume to decrease as the number of participating devices is reduced.
For full FSDP, the reduction is gradual since sharding occurs across all GPUs. 
In contrast, FSDP+TP exhibits a steeper reduction because FSDP operates across nodes, leading to a 33\% decrease in per-node communication volume when scaling from 4 to 2 nodes.
In contrast, TP+PP does not exhibit a similar reduction, as its inter-node communication is dominated by pipeline-parallel activation exchanges, which are point-to-point and remain largely invariant to the number of nodes.

\begin{figure}
\centering
\includegraphics[width=0.95\linewidth]{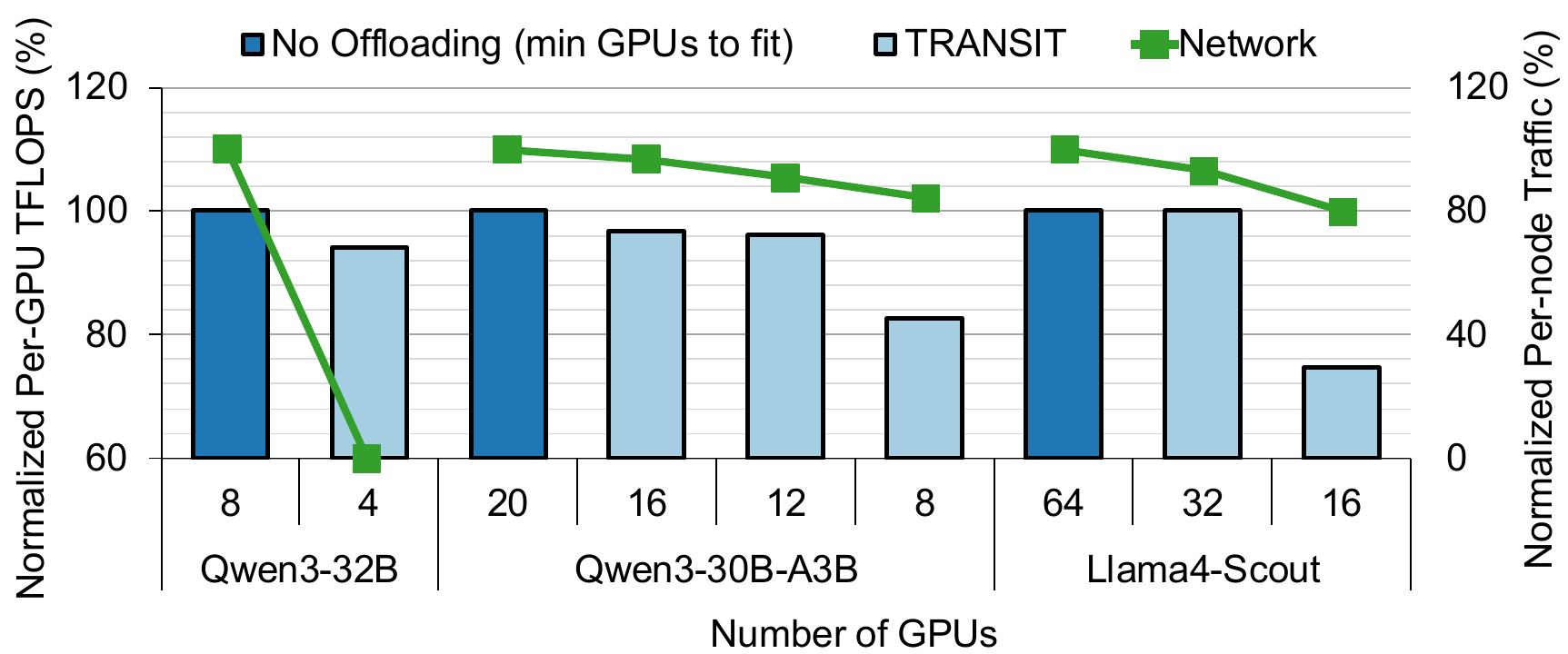}
\caption{Per-GPU TFLOPS of the no-offloading baseline and \framework\ under scale-in across models, along with per-node network traffic normalized to the baseline.}
\label{fig:result:different_model}
\end{figure}

\niparagraph{Different models and scales.}
\Cref{fig:result:different_model} illustrates the scaling behavior of different models across varying node counts on Cluster-A.
Despite differences in model size and minimum GPU requirements, a consistent trend emerges: \framework\ sustains per-GPU throughput comparable to the no-offloading baseline while using only half the number of GPUs.
These results show that TRANSIT preserves high per-GPU efficiency across the evaluated dense and MoE models despite differences in memory footprint and parallelization.
However, per-GPU TFLOPS decrease noticeably under aggressive scale-in.
For example, throughput drops by 18\% and 25\% for Qwen3-30B-A3B and Llama4-Scout, respectively, when the GPU count is reduced by 60\% and 75\%.
This degradation arises from higher memory oversubscription, which amplifies data movement overhead and limits the effectiveness of memory offloading.

\niparagraph{Implications for training efficiency.}
Scale-in trades aggregate training throughput for a smaller GPU allocation.
For a fixed training workload, aggregate throughput scales with the product of GPU count and per-GPU throughput, while GPU-hours are inversely proportional to per-GPU throughput.
Thus, preserving per-GPU throughput limits the increase in GPU-hours despite longer training times on fewer GPUs. 
When per-GPU throughput exceeds the baseline, scale-in can additionally reduce total GPU-hours.
These results highlight that \framework\ primarily enables memory-efficient GPU consolidation, with further resource savings when reduced communication overhead offsets CPU-memory access costs.

\subsection{Network-Constrained Training}
\label{sec:network}


\niparagraph{Bandwidth sensitivity.}
We quantify bandwidth sensitivity as $\rho=P(400)/P(100)$, where $P(B)$ is training throughput at per-NIC bandwidth $B$ (Gbps) for a fixed workload and GPU count.
A ratio closer to one indicates lower sensitivity to network bandwidth.

\begin{figure}
    \centering
    \includegraphics[width=\linewidth]{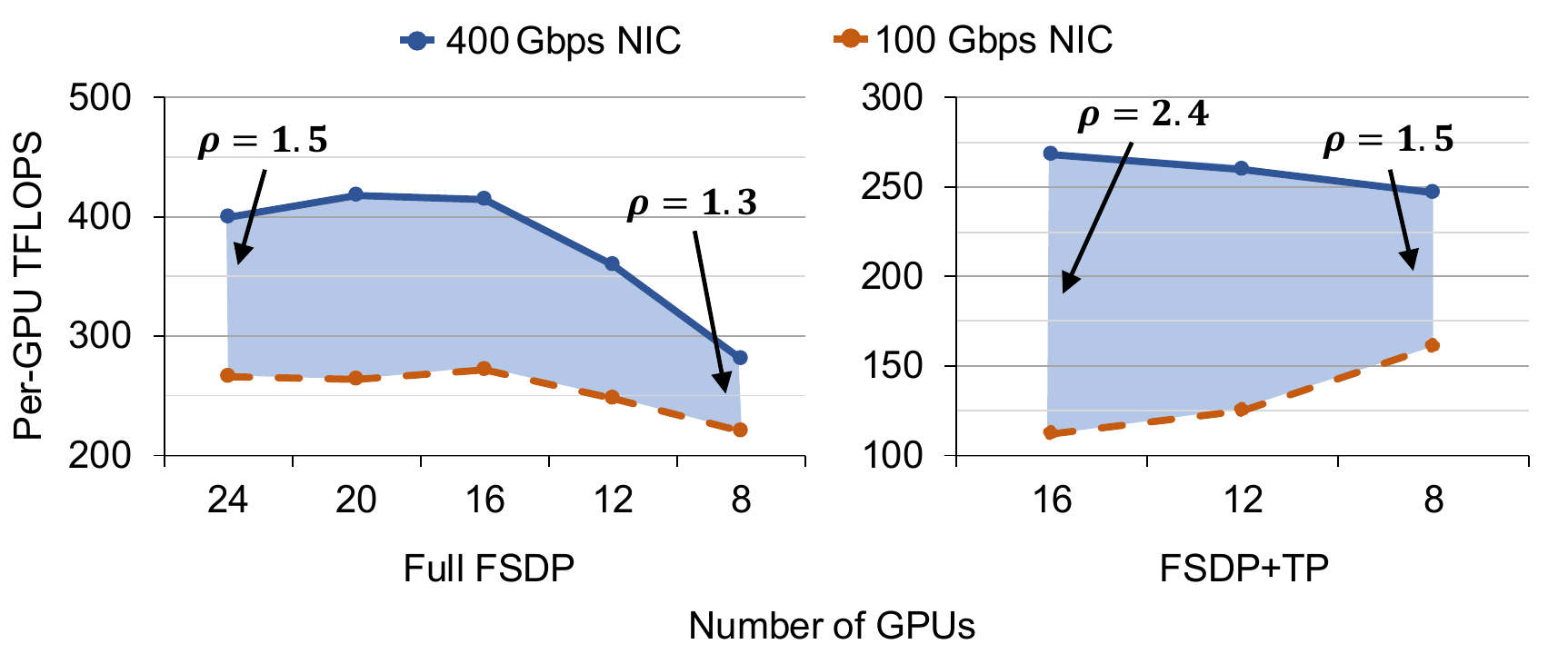}
    \caption{Per-GPU TFLOPS for Llama3-70B training across parallelism strategies and GPU counts under different NIC bandwidths (400\,Gbps vs. 100\,Gbps), along with the corresponding bandwidth sensitivity metric $\rho$.}
    \label{fig:result:different_nic_rate}
\end{figure}

Figure~\ref{fig:result:different_nic_rate} shows that scale-in with \framework\ reduces $\rho$ from 1.5 to 1.3 when scaling FSDP from 24 to 8 GPUs, and from 2.4 to 1.5 when scaling FSDP+TP from 16 to 8 GPUs.
Scale-in increases CPU-memory access overhead but reduces the communication component of iteration time, lowering the relative throughput penalty when NIC bandwidth drops from 400 to 100 Gbps.
Notably, at 100\,Gbps, eight-GPU \framework\ FSDP+TP achieves 161 per-GPU TFLOPS, 44\% higher than the 16-GPU no-offloading baseline (112 TFLOPS).
This demonstrates that reduced communication overhead can outweigh the additional CPU-memory access cost, enabling higher per-GPU efficiency with half the GPUs.

\subsection{Sensitivity Study}
\label{sec:sensitivity}

\begin{figure}[t]
    \centering
    \includegraphics[width=0.95\linewidth]{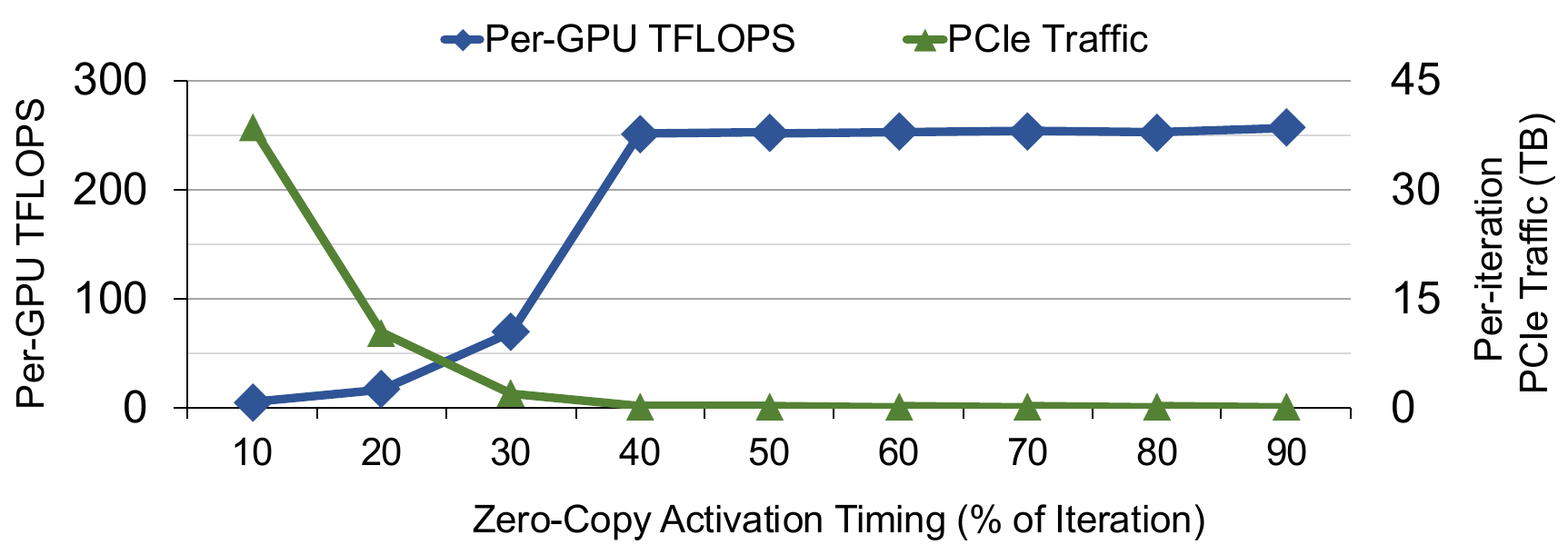}
    \caption{Sensitivity of Llama3-70B training throughput and per-iteration PCIe traffic to the zero-copy activation point.}
    \label{fig:advise_sensitivity}
\end{figure}

\niparagraph{Zero-copy activation timing.}
We evaluate the sensitivity of zero-copy activation timing using a representative Llama3-70B FSDP training rank with the model states and memory footprint corresponding to an eight-GPU configuration. 
We vary the activation point across an iteration, expressed as the fraction of completed kernel launches, and measure the resulting per-GPU training throughput.
\Cref{fig:advise_sensitivity} shows that early zero-copy activation severely degrades performance.
At 10\%, throughput drops to 4.8\,TFLOPS, with 38.6\,TB of PCIe traffic per iteration.
This occurs because optimizer updates in the preceding iteration can displace frequently accessed data, such as activations, gradients, and half-precision parameters, from HBM.
Enabling zero-copy too early prevents UVM from restoring their HBM residency, resulting in excessive remote accesses.
Delaying activation allows UVM to restore hot data to HBM while evicting less frequently accessed optimizer states to CPU memory.
Beyond 40\%, throughput remains stable at 251--257\,TFLOPS, with PCIe traffic below 0.23\,TB per iteration.
The broad favorable window supports using kernel-launch progress as a low-overhead activation heuristic, without identifying individual memory accesses or training phases.

\subsection{Cluster-level Simulation}
\niparagraph{Simulator design.} 
We build a discrete-event simulator using production traces to evaluate the cluster-wide impact of \framework.
The simulator models multi-rack clusters with 4 or 8 GPUs per node and 32 GPUs per rack, varying the number of racks to scale cluster capacity.
The scheduler prioritizes intra-rack placement using packing or spreading strategies~\cite{kub-pod, verma2015large}, falling back to multi-rack placement when necessary.

For cluster-level evaluation, we compare two scenarios: (1) a baseline that replays the original trace without modification and (2) a \framework-enabled scenario that reduces node allocations by 25--50\% for eligible jobs.
We leave single-node jobs unchanged and apply 25–50$\%$ node reductions for multi-node jobs, representing workloads whose memory-driven allocations permit scale-in.
Assuming unchanged per-GPU throughput, job duration initially scales inversely with GPU allocation (\eg halving the allocation doubles duration); placement-dependent throughput adjustments are then applied.
On top of this, the simulator distinguishes cases where cross-rack communication is eliminated, introduced, or partially reduced, and incorporates these effects into job duration.
The simulator incorporates scaling factors to capture these effects under different placement transitions, extracted from the results in \S\ref{sec:network}.
Specifically, when an inter-rack job becomes intra-rack, we apply a speedup factor derived from \framework\ at 400\,Gbps NIC configuration, normalized to a no-offloading baseline at 100\,Gbps NIC configuration (0--137\%).
When an intra-rack job becomes inter-rack, we apply a slowdown factor based on \framework\ at 100\,Gbps NIC configuration, normalized to a 400\,Gbps baseline NIC configuration (0--55\%).
For jobs that remain intra-rack or inter-rack before and after applying \framework, we apply a scaling factor corresponding to \framework\ performance relative to the baseline under the same placement and network setting (ranging from $-5\%$ to $5\%$).
The simulator reports a range of cluster-level metrics across both eligible jobs and all jobs, including cluster throughput, GPU-hours per job, queue time, and inter-rack traffic.

\niparagraph{Workload.} Input traces are preprocessed from the ACME datacenter trace~\cite{hu2024characterization}, specifically the Kalos and Seren subsets.
Kalos represents a relatively lightly loaded workload, while Seren captures a more heavily loaded setting.

\begin{figure}
    \centering
    \begin{subfigure}[b]{0.4776\linewidth}
        \centering
        \includegraphics[width=\textwidth]{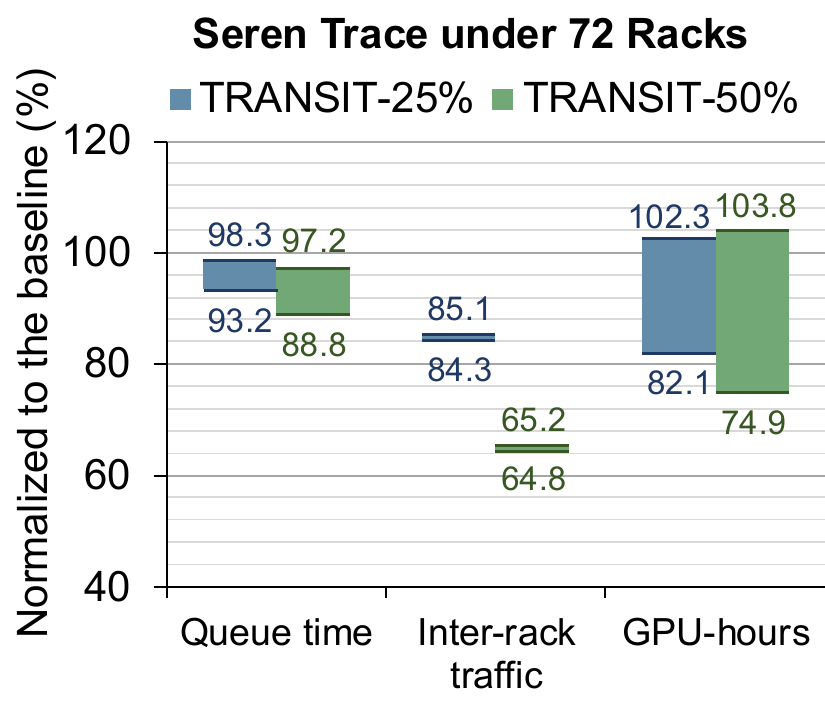}
        \vspace{-15pt}
        \caption{}
        \label{fig:result:cluster_full_scale}
    \end{subfigure}
    \hfill
    \begin{subfigure}[b]{0.5123\linewidth}
        \centering
        \includegraphics[width=\textwidth]{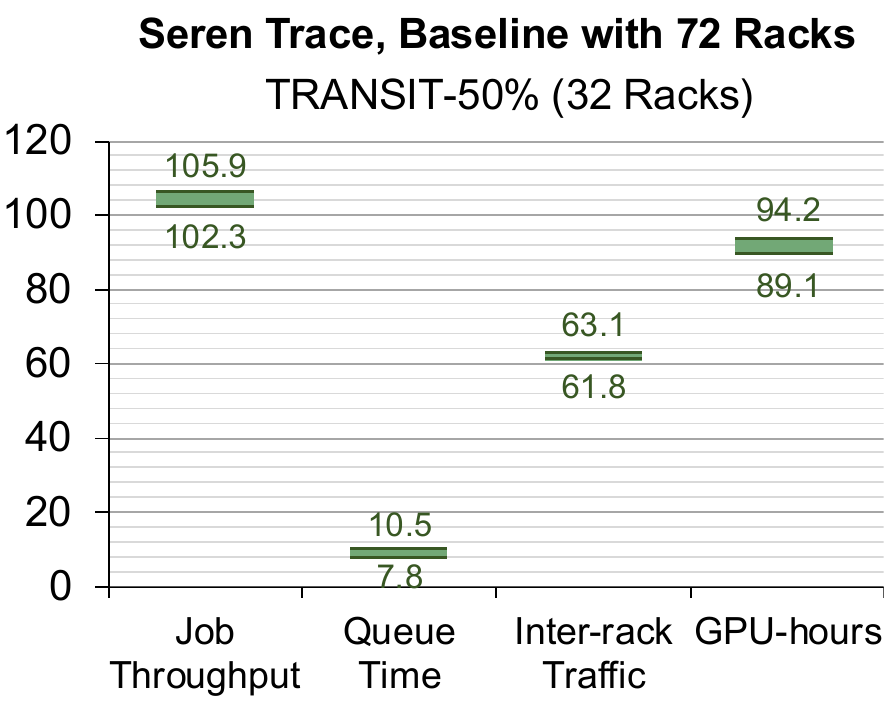}
        \label{fig:result:cluster_scaled}
        \vspace{-15pt}
        \caption{}
    \end{subfigure}
    \caption{Cluster-level impact of \framework\ on the Seren trace: (a) full-scale 72-rack cluster and (b) reduced-scale 32-rack \framework\ deployment compared against the 72-rack baseline. Bars show normalized results with ranges.}
    \label{fig:result:cluster}
\end{figure}

\niparagraph{Performance improvement across different loads.}
We begin with the Seren trace on a full-scale 72-rack cluster, representing a heavily loaded setting.
As shown in \Cref{fig:result:cluster_full_scale}, \framework\ reduces inter-rack traffic by 14.9--15.7\% and 34.8--35.2\% at 25\% and 50\% node count reductions, respectively, demonstrating that greater node consolidation consistently improves placement locality.
This improved locality also translates into system-level benefits, reducing average eligible-job queue time by 1.7--6.8\% and 2.8--11.2\%, respectively.
Moreover, \framework\ reduces average GPU-hours per job by up to 18.0\% and 25.1\% at 25\% and 50\% node count reductions, respectively.
Only under the worst-case combination of slowdowns from node count reduction and unfavorable inter-rack placement transitions do GPU-hours increase, by up to 2.3\% and 3.8\%, respectively.
In the 8 GPUs per node configuration, 40.6\% of eligible jobs collapse to a single node under 50\% node count reduction, eliminating inter-node communication for those jobs.

We next evaluate the Kalos trace on a full 76-rack cluster, representing a lightly loaded setting.
Here, cluster-level queue time benefits do not materialize because freed nodes remain idle rather than serving queued jobs.
Nevertheless, inter-rack traffic decreases by 24.2\% and 49.2\% under 25\% and 50\% node count reductions, respectively, while idle GPU time changes by less than 1\%.

\niparagraph{Performance under reduced cluster size.}
We next evaluate \framework\ under capacity constraints using the Seren trace on a reduced 32-rack cluster, less than half the 72-rack production deployment, and compare against a baseline running the same workload on the full 72-rack cluster without \framework.
As shown in \Cref{fig:result:cluster_scaled}, at 50\% node count reduction, \framework\ improves eligible-job throughput by 2.3--5.9\%, reduces average eligible-job queue time by 89.5--92.2\%, and decreases inter-rack traffic by 36.9--38.2\% across the evaluated throughput assumptions.
Furthermore, GPU-hours per job decrease by 5.8--10.9\%, showing that, under the modeled scale-in and placement assumptions, smaller GPU allocations can offset substantially reduced cluster capacity.

For the Kalos trace on a reduced 32-rack cluster, \framework\ reduces queue time by 85.2\% and 95.3\% at 25\% and 50\% node count reductions, respectively, relative to the full-scale baseline.
Notably, varying the assumed throughput improvement for jobs transitioning from inter-rack to intra-rack communication has negligible impact, indicating that the gains are driven primarily by node count reduction and improved placement locality rather than per-job throughput changes.

%% file: Sections/7_related.tex
\section{Related Work}
\niparagraph{Training LLMs with data offloading.}
Fine-grained systems such as G10~\cite{zhang2023g10}, TeraIO~\cite{NEURIPS2025_46863e01}, and SSDTrain~\cite{wu2025ssdtrain} exploit tensor lifetimes and execution characteristics to schedule data movement across GPU HBM, CPU DRAM, and NVMe SSDs.
While effective, they require tight framework integration, which becomes increasingly complex under multi-dimensional parallelism.
In contrast, ZeRO-Offload~\cite{ren2021zero}, ZeRO-Infinity~\cite{10.1145/3458817.3476205}, and TorchTitan~\cite{liang2025torchtitan} favor simpler, coarse-grained offloading of entire data classes (\eg parameters or optimizer states).
Application-transparent offloading via NVIDIA Unified Virtual Memory (UVM)~\cite{nvidia-uvm} decouples memory management from training frameworks.
However, on-demand migration can incur repeated page faults under memory pressure, motivating custom prefetching and placement policies~\cite{jung2023deepum, lin2025forest}.
Forest~\cite{lin2025forest} modifies the UVM driver to control page placement and migration, while DeepUM~\cite{jung2023deepum} introduces kernel-space components and runtime integration to intercept page faults and orchestrate prefetching.
More broadly, G10, TeraIO, DeepUM, and Forest exploit tensor lifetimes or kernel-access correlations to guide data movement and placement.
In contrast, \framework\ uses recurring kernel-execution patterns to guide zero-copy access while leaving page residency to the driver, without requiring kernel modifications.

\niparagraph{Communication optimizations for distributed LLM training.}
Prior work reduces communication overhead through compression, parallelism, topology-aware placement, and computation--communication overlap.
ZeRO++~\cite{wang2023zero++} reduces communication volume through low-precision transfers and hierarchical partitioning.
Pipeline parallelism replaces global synchronization with stage-wise communication, while improved schedules mitigate pipeline inefficiencies~\cite{harlap2018pipedream, narayanan2021efficient}.
Prior work improves distributed training efficiency through communication-aware scheduling and resource allocation~\cite{cao2024crux, 227623, peng2018optimus, zheng2022alpa}, while ByteScheduler~\cite{peng2019generic} overlaps communication with computation through priority-based scheduling.
These approaches optimize communication within a given GPU allocation but do not directly address GPU allocations driven by training-state memory requirements. 
In contrast, \framework\ reduces memory-driven GPU requirements, potentially reducing inter-node communication and network sensitivity.

%% file: Sections/8_conclusion.tex
\section{Conclusion}
\label{sec:conclusion}

Conventional distributed LLM training often scales out to meet GPU memory requirements, while memory oversubscription generally incurs substantial performance penalties. Our work challenges this tradeoff. We presented \framework, demonstrating that execution-guided zero-copy access can make driver-managed memory oversubscription efficient without explicit data placement, prefetching, or framework modifications. By relaxing the coupling between GPU allocation and memory capacity requirements, \framework\ enables aggressive scale-in while preserving high per-GPU computational efficiency. Our evaluation further shows that using fewer GPUs can reduce inter-node communication and, in production-trace-driven simulation, improve cluster-level scheduling efficiency. Together, these results show that memory oversubscription can be a practical means of improving distributed LLM training efficiency, rather than merely serving as a fallback for insufficient GPU memory.

%% file: Sections/9_Acknowledgement.tex
\section*{Acknowledgement}
This work was supported in part by grants from the IBM-Illinois Discovery Accelerator Institute (IIDAI) and the Institute of Information \& Communications Technology Planning \& Evaluation (IITP) grant funded by the Korean government (MSIT) (No. RS-2025-02214497, Development of low-level optimization program API technology for AI semiconductors).
The authors acknowledge the National Artificial Intelligence Research Resource (NAIRR) Pilot (NAIRR250426), the AI Alliance and resources from IBM, Red Hat and the Mass Open Cloud for contributing to this research result.
The authors thank Padmanabhan Santhanam (IBM Research) for his constant support during the NAIRR Pilot administrative process, and Kevin O’Connor (IBM Reserach), Jason Schlessman (Red Hat), Taj Salawu (Red Hat), Naved Ansari (MOC), and Hakan Saplakoglu (MOC), for their support in provisioning and maintaining the infrastructure used for large-scale GPU experiments.